\documentclass[a4paper,11pt]{article}
\pdfoutput=1 
\usepackage[skip=10pt plus1pt, indent=5pt]{parskip}
\usepackage{jcappub} 

\usepackage[T1]{fontenc} 
\usepackage{enumitem}

\usepackage[english]{babel}				

\usepackage{amsmath}
\usepackage{amssymb}
\usepackage{amsthm}
\theoremstyle{definition}
\newtheorem{defn}{Definition}[section]
\usepackage{array}
\usepackage{multirow}
\usepackage{mathtools}					
\usepackage{xcolor}						

\usepackage[colorlinks=true,urlcolor=blue,citecolor=blue,linkcolor=blue]{hyperref}

\newcommand{\dd}{{\rm d}}

\newcommand{\Mp}{M_\text{Pl}}

\newcommand\LC[1]{\mathring{#1}{}}
\newcommand\LCG{\LC{\Gamma}}
\newcommand\LCD{\LC{\nabla}}
\newcommand\LCbox{\LC{\square}}
\newcommand\LCR{\LC{R}}

\newcommand\LCten{\varepsilon}

\newcommand\axcurrent{J}
\newcommand\trcurrent{L}

\newcommand\jump[1]{[\![#1]\!]}
\newcommand\udis[1]{\underline{#1}{}}
\newcommand\disLCD{\mathring{\mathcal{D}}{}}
\newcommand\regpart{\varrho \,}
\newcommand\singpart[1]{\Delta^{{\scriptscriptstyle \!\!(#1)}} \!}
\newcommand\sSigma{{\scriptscriptstyle \Sigma}}
\newcommand\del{\delta^\sSigma}

\newcommand\szero{{\scriptscriptstyle (0)}}
\newcommand\sone{{\scriptscriptstyle (1)}}

\newcommand\tosigma[1]{\overline{#1}}

\newcommand{\myskip}{\hspace{10pt}}
\newcommand{\mybigskip}{\hspace{20pt}}

\title{Junction conditions in bi-scalar Poincar\'e Gauge gravity}

\author[1]{Adri\'an Casado-Turri\'on,}
\affiliation[1]{Departamento de F\'isica Te\'orica and Instituto IPARCOS, Universidad Complutense, 28040 Madrid, Spain}
\author[2,3]{\'Alvaro de la Cruz-Dombriz,}
\affiliation[2]{Departamento de F\'isica Fundamental, Universidad de Salamanca, E-37008 Salamanca, Spain}
\affiliation[3]{Cosmology and Gravity Group, Department of Mathematics and Applied Mathematics, University of Cape Town, Rondebosch 7700, Cape Town, South Africa}
\author[4]{Alejandro Jim\'enez-Cano}
\author[4]{and Francisco Jos\'{e} Maldonado Torralba}
\affiliation[4]{Laboratory of Theoretical Physics, Institute of Physics, University of Tartu, W. Ostwaldi 1, 50411 Tartu, Estonia\\$\,$}

\emailAdd{adricasa@ucm.es}
\emailAdd{alvaro.dombriz@usal.es}
\emailAdd{alejandro.jimenez.cano@ut.ee}
\emailAdd{fmaldo01@ucm.es}

\abstract{
In this work, we study the junction conditions of the ghost-free subclass of quadratic Poincaré Gauge gravity, which propagates one scalar and one pseudo-scalar. For this purpose, we revisit the theory of distributions and junction conditions in gravity, giving a novel insight to the subject by introducing a convenient notation to deal with regular and singular parts. Then, we apply this formalism to bi-scalar Poincaré Gauge gravity and study some paradigmatic cases. We compare our results with the existing literature and the well-known predictions of General Relativity. We find that monopole spin densities are admissible, whereas both thin shells and double layers are allowed for the energy-momentum. Such layers can be avoided by setting appropriate continuity conditions on the dynamic fields of the theory, as well as on the Ricci scalar of the full connection and the Holst pseudo-scalar.
}

\begin{document}

\maketitle

\section{Introduction}
\label{sec:intro}

In any physical theory, one can always build a theoretical set-up consisting of two systems separated by a boundary layer. As a paradigmatic example, in classical electromagnetic theory, a charged surface produces a discontinuity in the electric field \cite{Misner:1974qy}. In general, the relations between the matter content at the surface and the (dis)continuities of the fields are known as \emph{junction conditions}. 

In gravitational theories, the theoretical framework consists of two spacetimes separated by a hypersuface. Chronologically, the junction conditions in General Relativity (GR) were first studied by Lanczos \cite{https://doi.org/10.1002/andp.19243791403}. Subsequently, several studies formulated them for timelike, spacelike, null, and general hypersurfaces, in terms of the geometric properties of such hypersurfaces \cite{darmois1927equations,Lichnerowicz:107002,o1952jump,bel1967conditions,taub1980space,bonnor1981junction,clarke1987junction,barrabes1989singular,Mars:1993mj}. For the case in which the dividing hypersurface is allowed to possess a matter content, they were obtained for timelike hypersurfaces by Darmois and Israel \cite{darmois1927equations,Israel:1966rt}, hence the name of \emph{Darmois-Israel's junction conditions}. The cases of null and general hypersurfaces were later studied in \cite{barrabes1991thin,Mars:1993mj}.

When considering theories of gravity surpassing the usual GR scenario, it becomes instrumental to study how the junctions conditions change, since
the allowed matched solutions of the underlying gravitational theory may differ from their GR counterparts. Paradigmatic examples of such studies include those in $f(R)$ gravity, in either the so-called metric \cite{Deruelle:2007pt,Clifton:2012ry,Senovilla:2013vra,Casado-Turrion:2022xkl}, Palatini \cite{Vignolo:2018eco,Olmo:2020fri}, and hybrid \cite{Rosa:2021mln} formalisms, quadratic gravity \cite{Reina:2015gxa}, generalized scalar-tensor theories \cite{Padilla:2012ze}, extended teleparallel gravity \cite{delaCruz-Dombriz:2014zaa}, Einstein-Cartan theory \cite{Arkuszewski:1975fz}, and metric-affine gravity \cite{Macias:2002sr}. 

In this communication we shall consider a ghost-free subclass of Poincar\'e Gauge gravity theories (PG for short). The ideas behind PG were first laid out by Sciama \cite{Sciama1963} and Kibble \cite{Kibble1961}. As their name suggests, PG theories can be obtained by gauging the Poincaré group, with the Riemannian curvature being the field strength of the homogeneous Lorentz group and the torsion being the field strength of the translation group. For a review of these theories, we refer the interested reader to \cite{Blagojevic:2013xpa,Blagojevic2001,Obukhov2006b}. The aforementioned ghost-free subclass of PG propagates, apart from the usual graviton, one scalar and one pseudo-scalar. That is why we shall refer to it as bi-scalar Poincaré Gauge gravity along this article. It is also worth noticing that this theory can be thought of as a subclass of metric-affine gravity, and it has both metric and Palatini $R+R^2$ (without nonmetricity) as sub-cases.

As mentioned before, the junction conditions of generic metric-affine gravity (of which PG is a sub-class) were first studied in \cite{Macias:2002sr}. In spite of this, as seen below, the present study about the junction conditions brings several original conclusions which cannot be obtained from the results contained in that reference. This is because junction conditions in our present understanding (i.e., the relations between the matter content of the matching hypersurface and the discontinuities of the fields) were not present in \cite{Macias:2002sr}. For this reason, the studies in \cite{Senovilla:2013vra,Olmo:2020fri,Rosa:2021mln} will be useful for comparison of our results, as their treatment of junction conditions is closer to ours, compared with the one found in \cite{Macias:2002sr}.

This article is organized as follows. In Section \ref{sec:biscalar} we present the basic geometrical quantities involved in bi-scalar Poincaré Gauge gravity, as well as the action and the equations of motion for vanishing tensor part of the torsion. Next, Section \ref{sec:GeneralJunctions} is devoted to introducing the general mathematical formalism of tensor distributions and junction conditions: first, in Subsec.~\ref{sec:matching}, we present the geometrical setting of matching spacetimes; then, in Subsecs.~\ref{sec:defregtensor} and \ref{sec:singparts}, we explore the decomposition of tensor distributions into regular and singular parts; in Subsec.~\ref{sec:metricdistrib}, we apply all this theoretical formalism to the particular case of the metric and its associated objects; and, in Subsec.  \ref{sec:procedure}, we present the procedure to be followed in the upcoming sections in order to obtain the junction conditions. Thus, in Section \ref{sec:JC}, all the previous tools are employed to analyze bi-scalar Poincaré gravity: in Subsec.~\ref{sec:consistency}, we find the consistency conditions needed to avoid products of singular distributions; in Subsec.~\ref{sec:decomp} we identify the singular pieces of the equations of motion; 
in Subsec.~\ref{sec:partcases}, we study particular cases for specific values of the parameters and compare with the literature; and
in Subsec.~\ref{sec:Dirac} we discuss how to employ the previous results in the Dirac case (which requires the Vielbein formulation). Finally, in Section \ref{sec:conclusions}, we collect the conclusions of this communication.

Some useful relations regarding the derivatives of distributions are provided in Appendix \ref{app:regsing}. In Appendix \ref{app:EoM}, we give the field equations of the theory and comment the non-dynamical role of the tensor irreducible part of the torsion. Then, in Appendix \ref{app:subtleties}, we comment some pertinent subtleties encountered when dealing with products of quantities to be promoted to distributions. Finally, in Appendix \ref{app:tablesterms}, we resort to three tables
in order to display the different terms that appear in the field equations, indicating their decomposition into regular and singular pieces.

\paragraph*{\textbf{Notation and conventions.}}
 
Throughout this article, we use the signature $(-,+,...,+)$ for the Lorentzian  spacetime metric and units such that $c=1$ and $8\pi G=\Mp^{-2}$. We also introduce $H_{(\mu \nu)}\equiv \frac{1}{2!}(H_{\mu\nu}+H_{\nu\mu})$ and $H_{[\mu\nu]}\equiv \frac{1}{2!}(H_{\mu\nu}-H_{\nu\mu})$, and analogously for objects with $n$ indices instead of two. Greek indices ($\mu,\nu,\rho,\sigma,\lambda...$) refer to arbitrary coordinates in spacetime and run from $0$ to $\dim \mathcal{M}-1$, whereas lowercase Latin indices ($a,b,c...$) represent coordinate indices in the matching hypersurface $\Sigma$. Regarding uppercase Latin indices, $A,B,...$ will be coordinate multi-indices (abbreviations for an arbitrary set of upper and/or lower indices $\mu,\nu,\rho...$) and $I,J,...$ (only used in Section \ref{sec:Dirac}) indicate components in an arbitrary orthonormal frame. Einstein's summation convention is used throughout the work unless otherwise stated. 

For the connection and curvature tensors we use the conventions in Wald's book \cite{Wald:1984rg}, which we summarize here for completeness. The covariant derivative of a vector is given by $\nabla_\mu V^\rho=\partial_\mu V^\rho + \Gamma^{\rho}{}_{\mu\nu} V^\nu$; the commutator of covariant derivatives is $[\nabla_\mu, \nabla_\nu]V^\lambda= -R_{\mu\nu\rho}{}^\lambda V^\rho - T^\sigma{}_{\mu\nu} \nabla_\sigma V^\lambda$, which is consistent with \eqref{eq:Tordef}; finally, the Ricci tensor is obtained as $R_{\mu\nu}\equiv R_{\mu\lambda\rho}{}^\lambda$. Objects associated to the Levi-Civita connection will be represented as $\LCG,\LCR,\LCD...$, and we denote $\LCbox\equiv g^{\mu\nu}\LCD_\mu \LCD_\nu$.

Overlined vectors, $\tosigma{V}_\mu$, are fully projected vectors onto $\Sigma$, whereas the normal part is denoted $V_\perp$ (see \eqref{eq:Vdecomp}).

Whenever there is a tensor with a distributional counterpart, we will underline the distributional version for clarity.

\section{Bi-scalar Poincaré Gauge gravity}
\label{sec:biscalar}

In the framework of PG, in addition to the metric (equivalently, the coframe fields), a non-trivial connection appears as a new basic field of the theory. By non-trivial we mean that it does not coincide with the Levi-Civita connection of the metric. To be precise, the connection in PG is metric-compatible, namely, the metric is parallel with respect to it ($\nabla_\mu g_{\nu\rho}=0$), but it is allowed to have a non-trivial antisymmetric part in its two lower indices. In other words, we can always write the connection as:
\begin{equation}
    \Gamma^\sigma{}_{\mu\nu} = \LCG^\sigma{}_{\mu\nu} + K ^\sigma{}_{\mu\nu}\,,
\end{equation}
where $\LCG^\sigma{}_{\mu\nu}$ is the Levi-Civita connection of the metric, and $K^\sigma{}_{\mu\nu}$ is the so-called \emph{contorsion tensor}. The latter can be written as
\begin{equation}
    K^\sigma{}_{\mu\nu} \equiv \frac{1}{2} g^{\rho\sigma} (T_{\rho\mu\nu}+T_{\nu\rho\mu}-T_{\mu\nu\rho}),
\end{equation}
where we have introduced the \emph{torsion tensor},
\begin{equation}
    T^\sigma{}_{\mu\nu}\equiv 2\Gamma^\sigma{}_{[\mu\nu]}. \label{eq:Tordef}
\end{equation}
In general, the torsion can be decomposed as follows into irreducible pieces under the pseudo-orthogonal group \cite{McCrea:1992wa}:
\begin{equation}
    T^\sigma{}_{\mu\nu} = t^\sigma{}_{\mu\nu} + \frac{2}{3}T_{[\mu}\delta^\sigma_{\nu]} + \frac{1}{6} \LCten^{\mu\nu\sigma\lambda}S_\lambda\,,
\end{equation}
where $T_\mu = T^\sigma{}_{\mu\sigma}$ is the trace vector, $S_\mu= \LCten_{\mu\nu\sigma\lambda}T^{\nu\sigma\lambda}$ is the axial vector, and the tensor piece $t^\sigma{}_{\mu\nu}$ is traceless and has vanishing totally antisymmetric part, $t_{[\mu\nu\sigma]}=0$.

The curvature can be expanded as:
\begin{equation}
    R_{\mu\nu\rho}{}^\sigma = \LCR_{\mu\nu\rho}{}^\sigma - 2\LCD_{[\mu} K^{\sigma}{}_{\nu]\rho} - 2K^\sigma{}_{[\mu|\lambda} K^\lambda{}_{|\nu]\rho}.
\end{equation}
It is then straightforward to derive the corresponding post-Riemannian expansions of the Ricci scalar ($R\equiv R_\mu{}^\mu$) and the Holst pseudo-scalar ($\mathcal{H}\equiv\LCten^{\mu\nu\rho\sigma}R_{\mu\nu\rho\sigma}$):
\begin{align}
    R &= \LCR + 2 \LCD_\mu T^\mu +\frac{1}{24}S_\mu S^\mu -\frac{2}{3}T_\mu T^\mu +\frac{1}{2} t_{\mu\nu\rho}t^{\mu\nu\rho}\,, \label{eq:RicciPostRiem} \\
   \mathcal{H}   &= \frac{2}{3}S_\mu T^\mu -\LCD_\mu S^\mu+\frac{1}{2} \LCten_{\mu\nu\rho\sigma}t^{\lambda\mu\nu}t_\lambda{}^{\rho\sigma}\,. \label{eq:HolstPostRiem}
\end{align}

In this investigation, we will focus on the theory whose action is given by:
\begin{equation}
    \mathcal{S}[g^{\mu\nu}, T_\mu, S_\mu, t^\rho{}_{\mu\nu}, \Psi]=\mathcal{S}_\text{g}[g^{\mu\nu}, T_\mu, S_\mu, t^\rho{}_{\mu\nu}] + \mathcal{S}_\text{m}[g^{\mu\nu}, T_\mu, S_\mu, \Psi],\label{eq:fullaction}
\end{equation}
with 
\begin{equation}
    \mathcal{S}_\text{g} = \int \dd^4x \sqrt{|g|}\left( \frac{\Mp^2}{2} R + \beta R^2 + \alpha \mathcal{H}^2 + \frac{m_T^2}{2}T_\mu T^\mu + \frac{m_S^2}{2} S_\mu S^\mu + \frac{m_t^2}{2} t_{\mu\nu\rho} t^{\mu\nu\rho}\right), \label{eq:gravaction}
\end{equation}
where $m_T$, $m_S$ and $m_t$ are mass parameters and $\alpha$ and $\beta$ are dimensionless coupling constants. Observe that we are assuming that the matter sector, as given by $\mathcal{S}_\text{m}$, only feels the dynamical pieces of the torsion, i.e.~$S_\mu$ and $T_\mu$.\footnote{
    As shown in Appendix \ref{app:EoMt}, the tensor part of the torsion $t^\lambda{}_{\mu\nu}$ is non-dynamical in this theory. Of course, one could consider a scenario in which it becomes dynamical by introducing {\it ad hoc} non-minimal derivative couplings to the matter.
}
Even under this restriction, the matter action in \eqref{eq:fullaction} contains many interesting particular cases. For example, scalars or U(1) gauge fields, which are, in fact, independent of the torsion; as well as the Dirac Lagrangian minimally coupled to the torsionful connection, which depends solely on $S_\mu$. At this point, we highlight that, strictly speaking, the Dirac Lagrangian would not fit in \eqref{eq:fullaction}, as it requires the use of the Vielbein formulation; however, the application of our results in that case is straightforward, as explained in Section \ref{sec:Dirac}.

The theory given by $\mathcal{S}_\text{g}$ is known as \emph{bi-scalar Poincaré Gauge gravity}. As  mentioned in the Introduction, it is a sub-case of the general quadratic PG theory \cite{Obukhov2006b,Blagojevic2001} that does not propagate ghostly degrees of freedom \cite{BeltranJimenez:2019hrm}. It can be shown that, besides the graviton, two scalars propagate in this theory (at least generically). However, we will work in terms of the vector variables, since the matter variations will then be more closely connected with the notion of spin density (according to the standard gauge framework), facilitating the interpretation of results.

From now on, and for the reasons explained in Appendix \ref{app:EoMt}, we will focus on matching configurations with $t^\lambda{}_{\mu\nu}=0$. In particular, all objects depending on $t^\lambda{}_{\mu\nu}$ ($R, \mathcal{H}$...) will be assumed to be evaluated in $t^\lambda{}_{\mu\nu}=0$. As a consequence, the equation of motion of $t^\lambda{}_{\mu\nu}$ trivializes (see Appendix \ref{app:EoM} for more details) and the remaining ones, namely the equations of motion of $T_\mu$, $S_\mu$ and $g_{\mu\nu}$, are respectively given by
\begin{align}
    \dfrac{4\alpha}{3}\mathcal{H} S_\mu +\left(M_T^2-\dfrac{8\beta}{3}R\right)T_\mu - 4\beta\LCD_\mu R&= \trcurrent_\mu, \label{eq:EoM1}\\
     \dfrac{4\alpha}{3}\mathcal{H} T_\mu +\left(M_S^2+\dfrac{\beta}{6}R\right)S_\mu +2\alpha\LCD_\mu \mathcal{H}&=  \axcurrent_\mu, \label{eq:EoM2}\\
   E_{\mu\nu}-\frac{1}{2}g_{\mu\nu}E_\rho{}^\rho-4\beta\Mp^{-2} (\LCD_{\mu}\LCD_{\nu}- g_{\mu\nu}\LCbox) R &= \Mp^{-2} T_{\mu\nu} \,.\label{eq:EoM3}
\end{align}
Here, we have introduced the tensor $E_{\mu\nu}$, defined in \eqref{eq:deftensorE}, as well as the parameters
\begin{equation}
    M_T^2\equiv m_T^2-\frac{2\Mp^2}{3},\mybigskip M_S^2\equiv m_S^2 + \frac{\Mp^2}{24}, \label{eq:defMSMT}
\end{equation}
and the matter currents
\begin{equation}
    \trcurrent^\mu\equiv -\frac{1}{\sqrt{|g|}}\frac{\delta \mathcal{S}_\text{m}}{\delta T_\mu},\mybigskip \axcurrent^\mu\equiv -\frac{1}{\sqrt{|g|}}\frac{\delta \mathcal{S}_\text{m}}{\delta S_\mu},\mybigskip T_{\mu\nu}\equiv -\frac{2}{\sqrt{|g|}}\frac{\delta \mathcal{S}_\text{m}}{\delta g^{\mu\nu}}\,, \label{eq:defLJT}
\end{equation}
which are respectively the vector spin density, the axial vector spin density and the energy-momentum tensor. Observe that the axial current $\axcurrent^\mu$ is indeed a pseudo-vector (it gets an extra sign under a reflection of all the coordinates), as can be easily deduced by consistency of the equation of motion of $S_\mu$,  \eqref{eq:EoM2}.

\section{Mathematical framework}
\label{sec:GeneralJunctions}

\subsection{The matching hypersurface}
\label{sec:matching}

For the discussion herein, we shall closely follow the approach in \cite{Reina:2015gxa}. Thus, let us consider two smooth Lorentzian manifolds  $(\mathcal{V}^\pm, g^\pm)$ with boundaries  $\Sigma^\pm$, such that: (i) $\Sigma^\pm$ is a smooth embedded submanifold in $\mathcal{V}^\pm$ of codimension 1, (ii) they are both either timelike or spacelike, and (iii) $\Sigma^+$ and $\Sigma^-$ are diffeomorphic. Then we construct the smooth manifold $\mathcal{M}$ resulting from gluing $\mathcal{V}^\pm$ along $\Sigma^\pm$, by using a suitable diffeomorphism (that we assume to be known). As a result, we get a submanifold $\Sigma$ of $\mathcal{M}$ separating the ``plus'' and the ``minus'' regions that we denote $\mathcal{M}^\pm$ (they exclude the hypersurface), so
\begin{equation}
    \mathcal{M} = \mathcal{M}^+ \sqcup \mathcal{M}^- \sqcup \Sigma \,. 
\end{equation}

Since $\mathcal{M}$ is a smooth manifold, one could take a chart around any given point $p$ in the matching hypersurface $\Sigma$, which intersects both sides of it. We call such a chart $\{x^\mu\}$, and, hereafter, all tensor components will be referred to it. This simplifies the formalism, although, in practice, one normally starts with different coordinate charts on both sides and one should do some preliminary work to connect them via parametric equations (see e.g.~\cite[Sec. 3.8]{poisson2004relativist}).

At this point, one can use all of the machinery of the theory of submanifolds around $\Sigma$ from both sides. We will work with a generic chart $\{y^a\}$ in $\Sigma$. We introduce, as usual, a basis of vectors $\{n\equiv n^\mu \partial_\mu, e_a \equiv \frac{\partial x^\mu}{\partial y^a}\partial_\mu\}$ for the tangent space of any point in $\Sigma$, which is the same from both sides (i.e., there is no distinction between $\{n^{+}, e^+_a\}$ and $\{n^{-}, e^-_a\}$), and where $n^\mu$ is the \emph{unit normal vector} of the hypersurface, chosen to point from $\mathcal{M}^-$ to  $\mathcal{M}^+$. As mentioned before,  $\Sigma$ is assumed to be either purely spacelike or timelike, which is thus characterized by the quantity
\begin{equation}
   \epsilon \equiv n^\mu n_\mu = \pm 1\,.  \label{eq:defepsilon}
\end{equation}
The \emph{induced metrics} from both sides are given by
\begin{equation}
    h^\pm_{ab} \equiv\frac{\partial x^\mu}{\partial y^a}\frac{\partial x^\nu}{\partial y^b} g^\pm_{\mu\nu}\,.
\end{equation}
Notice that this implies $h^+_{ab}=h^-_{ab}$ if and only if the metric is continuous ($g^+_{\mu\nu}(p)=g^-_{\mu\nu}(p), \forall p\in\Sigma$). As we will comment below in Section \ref{sec:metricdistrib}, this is necessary in order to have a well-defined curvature in the distributional sense (see also e.g. \cite{Reina:2015gxa}). In practice we will assume that the metric (and, hence, the induced metric as well) is continuous, so we drop the $\pm$ from the symbol.

Finally, we introduce two relevant symmetric tensors defined on $\Sigma$ that are purely transversal, i.e., such that their contractions with the normal vector are vanishing. We define the \emph{projector} as
\begin{equation}
    h_{\mu\nu} \equiv g_{\mu\nu}- \epsilon n_\mu n_\nu\,,
\end{equation}
and the \emph{extrinsic curvature} as
\begin{equation}
    K^\pm_{\mu\nu} \equiv h^\rho{}_\mu h^\lambda{}_\nu \LCD^\pm_\rho n_\lambda\,.   
\end{equation}
We will represent the trace simply as $ K^\pm \equiv g^{\mu\nu} K^\pm_{\mu\nu}$. Notice that the notation $\LCD^\pm$ is needed because, even though the metric is continuous, the corresponding Levi-Civita connection may have a non-vanishing discontinuity coming from the discontinuity of $\partial_\rho g_{\mu\nu}$.

We conclude this part by reminding that any 1-form field on $\Sigma$ can be decomposed into normal and tangential components as follows:
\begin{equation}
    V_\mu = \epsilon n_\mu V_\perp + \tosigma{V}_\mu,\label{eq:Vdecomp}
\end{equation}
where we have introduced the notation $V_\perp \equiv n^\mu V_\mu $ and $\tosigma{V}_\mu \equiv h^\lambda{}_\mu V_\lambda$. 

\subsection{Regular tensor distributions}
\label{sec:defregtensor}

For an introduction to the basic ideas about distributions on manifolds, we refer the reader to  Appendix A in \cite{Reina:2015gxa}. Here, we shall only concentrate on the definitions and nomenclature that are relevant for our purposes.

In general, a tensor distribution can be associated to any locally integrable tensor field. In our analysis, we will focus on a particular case of these, which will be refered to as ``regular tensor fields'' for convenience:

~

\begin{defn}\label{def:regularten}
A tensor field is said to be \emph{regular} if it satisfies the following requirements:
\begin{enumerate}[label=(\roman*)]
    \item It is smooth in $\mathcal{M}^{\pm}$.
    \item The tensor field and its successive (Levi-Civita) covariant derivatives have well-defined limits on the matching hypersurface from both sides.
\end{enumerate}
We notice that such limits could be non-coincident, i.e., the field and/or its derivatives could be discontinuous.
\end{defn}

~

Observe that, when doing the matching in a physical situation, there is an ambiguity in the value of the regular field right on the matching hypersurface. By convention, we \emph{choose} it to be the average of the limits from both sides of $\Sigma$.\footnote{
	Observe that this is a choice for ``basic'' tensors. Once we start combining them, the property that the value on $\Sigma$ coincides with this average (defined in \eqref{eq:defonsigma}) is not maintained. Indeed, by virtue of \eqref{eq:ssigmaproduct}, it is clear that given two fields fulfilling $A|_\Sigma = A^\sSigma$ and $B|_\Sigma=B^\sSigma$, the restriction of their product to $\Sigma$, 
    \[
    (AB)|_\Sigma=A|_\Sigma B|_\Sigma=A^\sSigma B^\sSigma\,,
    \]
    does not coincide with the average of the product coming from both sides, $(AB)^\sSigma = (A^+B^+ + A^-B^-)/2$. This is only true if either $A$ or $B$ is continuous.} 
Namely, if we introduce the Heaviside function \cite{abramowitz1972handbook},
\begin{equation}
    \Theta^\pm(p) \equiv \begin{cases} 1 & \text{if }p\in \mathcal{M}^\pm  \\  \frac{1}{2} & \text{if }p\in\Sigma \\ 0 & \text{if }p\in \mathcal{M}^\mp  \end{cases}
\end{equation}
then we can write a given regular tensor $T_A$ (where $A$ is an abbreviation for arbitrary tensor indices) as
\begin{equation}
    T_A = T^{+}_A \Theta^{+} + T^{-}_A \Theta^{-}.
\end{equation}
If we promote $\Theta^\pm$ to distributions $\udis{\Theta}^\pm$, we can associate a distribution to  $T_A$  as follows:
\begin{equation}
    T_A \mybigskip\to \mybigskip \udis{T}_A \equiv T^{+}_A \udis{\Theta}^{+} + T^{-}_A \udis{\Theta}^{-}\,.
\end{equation}
Distributions of this form (i.e.~those containing only terms with the Heaviside distribution) shall be named \emph{regular} as well.

Note that the underline is utilized just to distinguish between a given tensor and its distributional counterpart. It does not necessarily mean that the distribution is regular. For example, the Levi-Civita Riemann tensor $\LCR_{\mu\nu\rho}{}^\lambda$ has an associated distribution, to be represented as $\udis{\LCR}_{\mu\nu\rho}{}^\lambda$, which contains a singular part in general, as we will discuss later in Section \ref{sec:metricdistrib}.

For a given $p\in\Sigma$, we introduce
\begin{equation}
    T^\sSigma_A(p) \equiv \frac{1}{2}\big(T^+_A(p^+)+ T^-_A(p^-)\big)\,,\label{eq:defonsigma}
\end{equation}
where $T^\pm_A(p^\pm)$ is just an abbreviation for ``the limit of $T^\pm_A$ as we approach $p$ from the side $\mathcal{M}^\pm$''. The \emph{discontinuity} (also known as ``jump'') of the fields will be represented as:
\begin{equation}
    \jump{T_A}(p) \equiv T^+_A(p^+)- T^-_A(p^-) \,.\label{eq:defjump}
\end{equation}
We highlight that quantities $T^\sSigma_A$ and $\jump{T_A}$ are only defined on $\Sigma$. 

For objects defined on the hypersurface, such as $n^\mu$, $h^\pm_{\mu\nu}$ or $K^\pm_{\mu\nu}$, the quantity  \eqref{eq:defonsigma} and the discontinuity \eqref{eq:defjump} are not well-defined, since these tensors are not defined outside $\Sigma$. However, it is convenient to introduce the notation:\footnote{
    Notice that we can make these tensors compatible with the definitions \eqref{eq:defonsigma} and \eqref{eq:defjump} by extending them outside $\Sigma$ in an appropriate way. To do so, one can introduce a foliation given by the level sets of a certain scalar function $\Phi$, and such that $\Sigma = \{ p\in\mathcal{M} | \Phi=0\}$. See e.g. \cite[Sec. 3.7]{poisson2004relativist}.
    }
\begin{align}
    \jump{K_{\mu\nu}} &\equiv K^+_{\mu\nu} - K^-_{\mu\nu}\,,\\
    K^\sSigma_{\mu\nu} &\equiv \frac{1}{2} (K^+_{\mu\nu} + K^-_{\mu\nu})\,,
\end{align}
and similarly for $n^\mu$, $h_{\mu\nu}$, etc. Observe that, by construction,
\begin{equation}
    \jump{n^\mu}=0 \mybigskip \text{and}\mybigskip \jump{g_{\mu\nu}}=0~\Leftrightarrow~\jump{h_{ab}}=0\,.
\end{equation}

Given two regular tensors $A$ and $B$ (we drop the indices), it can be shown that
\begin{align}
    (AB)^\sSigma &= A^\sSigma B^\sSigma +\frac{1}{4}\jump{A}\jump{B} \,, \label{eq:ssigmaproduct}\\
    \jump{AB} &= A^\sSigma \jump{B} + \jump{A} B^\sSigma \,.
\end{align}
For the derivative of a regular scalar and a regular vector we find:\footnote{See \cite[App. D.2]{Reina:2015gxa} for the derivation of \eqref{eq:jumpDV} for $\epsilon=+1$ (i.e., for timelike $\Sigma$).}
\begin{align}
   \jump{\LCD_\nu f} &= \jump{\partial_\nu f} = \partial_\nu\jump{f}\,, \label{eq:jumpDS}\\
    \jump{\LCD_\nu V_\rho} &= \epsilon n_\nu n^\lambda \jump{\LCD_\lambda V_\rho} + h^\lambda{}_\nu \LCD^\sSigma_\lambda \jump{V_\rho} +\epsilon V^\sSigma_\lambda \left(\jump{K_{\nu\rho}}n^\lambda - \jump{K_\nu{}^\lambda}n_\rho\right)\,,\label{eq:jumpDV}
\end{align}
where $\LCD^\sSigma$ is the covariant derivative taken with respect to $\LCG^{\sSigma \rho}{}_{\mu\nu}$.

As an application of \eqref{eq:jumpDV}, it can be proved that (for vanishing $t^\rho{}_{\mu\nu}$):
\begin{align}
    \jump{R}&= \jump{\LCR} +2\epsilon n^\mu n^\nu \jump{\LCD_\mu T_\nu} +2 h^{\mu\nu}\LCD^\sSigma_\mu \jump{T_\nu}+2\epsilon T^\sSigma_\perp \jump{K} +\frac{1}{24}\jump{S_\mu S^\mu} -\frac{2}{3}\jump{T_\mu T^\mu} \,,\label{eq:jumpR}\\
    \jump{\mathcal{H}}&= \frac{2}{3}\jump{S_\mu T^\mu} - \epsilon n^\mu n^\nu \jump{\LCD_\mu S_\nu} -h^{\mu\nu}\LCD^\sSigma_\mu \jump{S_\nu}-\epsilon S^\sSigma_\perp \jump{K}\,. \label{eq:jumpH}
\end{align}

Before moving on with the singular parts we make two important remarks:
\begin{itemize}
    \item Firstly, we present for completeness the definition of the distributional derivative $\disLCD$ associated to the Levi-Civita connection $\LCD$. It is defined as the operator acting on a tensor-valued distribution $\udis{F}_A$ such that, when acting on a test tensor $\varphi^{\mu A}$ gives:
\begin{equation}
    \langle\disLCD_\mu\udis{F}_A, \varphi^{\mu A} \rangle \equiv \langle\udis{F}_A, -\LCD_\mu\varphi^{\mu A} \rangle\,.\label{eq:defdisD}
\end{equation}
    If the distribution $\udis{F}_A$ is regular and $F_A$ corresponds to its associated regular tensor, then: 
\begin{equation}
    \langle\disLCD_\mu\udis{F}_A, \varphi^{\mu A} \rangle = - \int F_A\LCD_\mu\varphi^{\mu A} \, \sqrt{|g|} \dd^4 x \,.
\end{equation}
In particular, $\disLCD_\mu \udis{\Theta}^\pm = \pm \epsilon n_\mu \del$\,. 
    \item Secondly, we recall that it is not possible to define a product of distributions in general. This is one of the main obstacles and the source of many ambiguities when developing a framework for junction conditions. We will bypass this problem by imposing appropriate consistency conditions and by choosing a suitable prescription when promoting products of tensors to the distributional framework. 
\end{itemize}

\subsection{Singular parts}\label{sec:singparts}

As we will justify below, all of the objects that constitute our equations of motion \eqref{eq:EoM1}-\eqref{eq:EoM3} can be thought of as distributions with the general form:
\begin{equation}
    \udis{F}_A  = 
    \regpart\left[\udis{F}_A\right]+ \singpart{0}\left[\udis{F}_A\right] +\singpart{1}\left[\udis{F}_A\right]+ \singpart{2}\left[\udis{F}_A\right]+...+ \singpart{N}\left[\udis{F}_A\right]\,, \label{eq:decompF}
\end{equation}
for some finite $N$. Here, $\regpart\left[\udis{F}_A\right]$ represents the \emph{regular part}, namely the one that can be written as
\begin{equation}
    \regpart\left[\udis{F}_A\right] =  F^{+}_A \, \udis{\Theta}^{+} +  F^{-}_A \,\udis{\Theta}^{-}\,,
\end{equation}
for a certain regular tensor field $F_A$. The rest of the terms in \eqref{eq:decompF} constitute the \emph{singular part} of $\udis{F}_A$, whose different pieces (monopolar, dipolar, quadrupolar...) follow the structure:
\begin{align}
   \singpart{0}\left[\udis{F}_A\right]&=  F^\szero_A \del \,,\label{eq:defD0}\\
   \singpart{1}\left[\udis{F}_A\right]&=  \disLCD_\mu \left( F^\sone_A n^\mu \del\right) \,,\label{eq:defD1}\\
   \singpart{2}\left[\udis{F}_A\right]&=  \disLCD_\mu \disLCD_\nu\left( F^{{\scriptscriptstyle(2)}}_A n^\mu n^\nu \del\right) \,,\\
   \vdots\nonumber\\
   \singpart{N}\left[\udis{F}_A\right]&=  \disLCD_{\mu_1}...\disLCD_{\mu_N}\left( F^{{\scriptscriptstyle(N)}}_A n^{\mu_1}... n^{\mu_N} \del\right) \,,\label{eq:defDN}
\end{align}
for certain tensors $\{ F^{{\scriptscriptstyle(\ell)}}_A\}^N_{\ell =0}$ defined on the matching surface, which can be interpreted as multipole densities of order $\ell$ \cite{Senovilla:2014yea}. To be precise, the objects $\{F^{{\scriptscriptstyle(\ell)}}_A\}^N_{\ell=0}$ are just certain regular tensor fields that are vanishing outside $\Sigma$ and take some finite value in it (e.g., the discontinuities $\jump{F_A}, \jump{\nabla_\mu F_A}$, ...). In Appendix \ref{app:regsing}, we collect useful expressions for some particular cases. 

As the expressions \eqref{eq:defD0}-\eqref{eq:defDN} suggest, singular parts of higher orders are indeed generated when taking successive distributional derivatives (Levi-Civita is assumed). Schematically:
\begin{equation}
    \regpart\, (\text{continuous}) \myskip \overset{\disLCD}{\longrightarrow}\myskip
    \regpart \myskip \overset{\disLCD}{\longrightarrow}\myskip
    \regpart + \singpart{0} \myskip \overset{\disLCD}{\longrightarrow}\myskip
    \regpart + \singpart{0}+ \singpart{1} \myskip \overset{\disLCD}{\longrightarrow}\myskip \ldots \label{eq:schemeD}
\end{equation}
Namely, if we start with a continuous distribution (with non-necessarily continuous derivative) we will first generate a regular piece that is discontinuous, then we start generating $\singpart{0}$, then $\singpart{1}$, and so on. With these ideas in mind, it is not difficult to see why we have introduced the general form \eqref{eq:decompF}. At the end of the day, the basic fields of the theory will be assumed to be regular (see Section \ref{sec:preconsistency}) and, since the equations of motion consist of combinations of these objects and their derivatives, only distributions of the type \eqref{eq:decompF} will appear in our computations.

\subsection{Geometrical quantities as distributions}
\label{sec:metricdistrib}

For a given metric, either a discontinuous regular part or a non-vanishing singular part would make the distributional Riemann tensor (and hence its contractions) ill-defined. Let us explain this in more detail. One simple way to understand this is by recalling that the curvature contains quadratic terms in the Christoffel symbols, thus being of the form
$\LCG\LCG~\sim~(g^{-1} \partial g)(g^{-1} \partial g)$. If the metric contains a singular part, then the Christoffel symbols are quadratic in singular pieces, and consequently ill-defined in the distributional sense. On the other hand, if the metric is regular but discontinuous, then the partial derivatives of the metric are singular and, since the Riemann is quadratic in them, we get that the curvature is ill-defined also in this case.  Then, the conclusion is that the continuity of the metric is needed for any gravitational theory with equations of motion depending on the Riemann tensor (or its traces); otherwise, $\LCR_{\mu\nu\rho}{}^\lambda$ would not be well-defined in the sense of distributions. In fact, from now on, we will simply write $g_{\mu\nu}$ instead of $\udis{g}_{\mu\nu}$, which is consistent with our prescription for products, given by \eqref{eq:prescTheDel}.\footnote{Observe that $\disLCD$ is also metric-compatible in the sense that $g_{\mu\nu} \disLCD_\rho \udis{F}_A =  \disLCD_\rho (g_{\mu\nu}\udis{F}_A)$, as can be derived from \eqref{eq:defdisD}.}

Nonetheless, notice that, even if the metric is continuous, the Riemann is an intrinsically quadratic object, so it will still involve products of distributions of the types $\udis{\Theta}\, \udis{\Theta}$ and $\udis{\Theta}\del$, however, as we will comment in the next section, we can make sense of these combinations by a suitable prescription (precisely that on equation \eqref{eq:prescTheDel}).

After these clarifications, we introduce the following distributions associated to a continuous metric:
\begin{align}
    \udis{\LCG}^\lambda{}_{\mu\nu} &\equiv \Gamma^{+ \lambda}{}_{\mu\nu} \udis{\Theta}^+ + \Gamma^{- \lambda}{}_{\mu\nu} \udis{\Theta}^- \,,\label{eq:distribGamma}\\
    \udis{\LCR}_{\mu\nu\rho}{}^\lambda &\equiv 
    \LCR^+_{\mu\nu\rho}{}^\lambda \udis{\Theta}^+ + \LCR^-_{\mu\nu\rho}{}^\lambda \udis{\Theta}^-  + \LCR^\szero_{\mu\nu\rho}{}^\lambda \del \,,\label{eq:distribRiemann}
\end{align}
with
\begin{equation}
    \LCR^\szero_{\mu\nu\rho}{}^\lambda \equiv -2\epsilon n_{[\mu} \jump{\LCG^\lambda{}_{\nu]\rho}} = 4g^{\lambda\sigma} n_{[\mu}\jump{K_{\nu][\rho}}n_{\sigma]}\,,\label{eq:singpart0Riem}
\end{equation}
where we are using the notation in \eqref{eq:defD0}, i.e., we have introduced $\singpart{0}\big[\udis{\LCR}_{\mu\nu\rho}{}^\lambda\big]\equiv \LCR^\szero_{\mu\nu\rho}{}^\lambda \del$.

Indeed, \eqref{eq:distribRiemann} is what we would have obtained from the usual definition of the Riemann tensor after substituting $\LCG \to \udis{\LCG}$ (as given in \eqref{eq:distribGamma}), and the partial derivatives by partial distributional derivatives.

From \eqref{eq:singpart0Riem}, the singular parts of the Ricci tensor distribution ($\udis{\LCR}_{\mu\nu} \equiv \udis{\LCR}_{\mu\rho\nu}{}^\rho$) and the Ricci scalar distribution ($\udis{\LCR} \equiv g^{\rho\sigma}\udis{\LCR}_{\rho\sigma}$), which are purely monopolar, are given by:
\begin{align}
    \LCR^\szero_{\mu\nu} & = - n_\mu n_\nu \jump{K}- \epsilon \jump{K_{\mu\nu}}\,,\\
    \LCR^\szero & = - 2\epsilon \jump{K}\,.
\end{align}

\subsection{Junction conditions: general procedure}
\label{sec:procedure}

Equipped with the previous formalism, and once a theory in curved spacetime, which depends on a metric and a certain family of fields, is given, we can stick to the following steps in order to obtain the junction conditions matching two manifolds as described in 
Section~\ref{sec:matching}:

~

\begin{enumerate}[label=(\roman*)]
\itemsep=10pt
    \item Compute the equations of motion of the theory.
    \item Write the equations of motion explicitly in terms of the basic fields of the theory (those appearing in the functional dependency of the action), and their Levi-Civita covariant derivatives. For the case of the metric, we will get the associated Riemann tensor, its traces and potential Levi-Civita covariant derivatives of them.
    \item Expand all the derivatives, i.e., use the Leibniz rule of the Levi-Civita derivative. We introduce this step in order to avoid facing distributional derivatives of products of distributions.
    \item Promote the metric and the rest of the fields to distributions of the type \eqref{eq:decompF}. Similarly, the covariant derivatives are promoted to distributional derivatives, i.e. $\LCD \to \disLCD$.
    \item Find a set of conditions (as minimal as possible) to avoid ill-defined terms in the sense of distributions in the equations of motion, e.g. $\sim \del\del$.

    ~
    
    Indeed, this is a key stage even when there are no quadratic terms in singularities, because, since the equations are non-linear, there are still products of the type $\udis{\Theta}\,\udis{\Theta}$ and $\udis{\Theta}\del$. An example of this is the product of $n$ regular fields, $X_i$, and an additional one, $Y$, which contains both regular and monopolar singular parts. If we promote to the distributional framework naively, we find 
    \begin{equation}
        X_1\ldots X_n Y \quad \overset{\text{naive}}{\longrightarrow} \quad \regpart[X_1] \ldots  \regpart[X_n](\regpart [Y] + \singpart{0}[Y])\,.
    \end{equation}
    However, it is possible to make sense of the products $\udis{\Theta}\,\udis{\Theta}$ and $\udis{\Theta}\del$ by introducing a consistent prescription. We define the promotion of these terms to the distributional framework as follows:
    \begin{align}
        X_1\ldots X_n Y \quad \overset{\text{def}}{\longrightarrow}  \quad \udis{X}_1 \ldots\udis{X}_n\udis{Y} &\equiv X^+_1\ldots X^+_n Y^+ \udis{\Theta}^+ \nonumber\\
        &\quad+ X^-_1\ldots X^-_nY^-\udis{\Theta}^-  \nonumber\\
        &\quad+ X^\sSigma_1 \ldots X^\sSigma_n \singpart{0}[Y]\,. \label{eq:prescTheDel}
    \end{align}
    For our purposes, only the cases $n=1$ and $n=2$ of this formula will be relevant. The result is the same as in the naive promotion after using appropriate identifications, needed to fix a certain ambiguity. A more detailed discussion on these subtleties can be found in Appendix \ref{app:subtleties}.
    
    \item Finally, under the restrictions pointed out in the previous step, one can perform a decomposition of the equations of motion into the different singular pieces as in \eqref{eq:decompF}, and find the conditions that the matter sources must fulfill in order to obtain the desired kind of matching, either smooth (with no surface sources) or non-smooth (including surface sources such as thin shells, double layers, etc.). The resulting conditions set important limitations on which manifolds can be properly joined.
\end{enumerate}

~

This procedure, as outlined above, is general for any gravitational theory. In the following section, we will apply it to the theory \eqref{eq:fullaction}-\eqref{eq:gravaction} described in Section \ref{sec:biscalar}.

\section{Junction conditions in bi-scalar Poincaré Gravity}
\label{sec:JC}

Given the equations of motion of the theory, \eqref{eq:EoM1}-\eqref{eq:EoM3}, the next step, according to the procedure above, is to write them in terms of the basic fields $\{g_{\mu\nu},T_\mu,S_\mu\}$ and their Levi-Civita covariant derivatives. To be precise, we should expand $R$ and $\mathcal{H}$ by using \eqref{eq:RicciPostRiem}-\eqref{eq:HolstPostRiem} and use the Leibniz rule  in the terms containing derivatives of $R$ and $\mathcal{H}$.

\subsection{Eliminating ill-defined terms from the gravitational sector}
\label{sec:consistency}

\subsubsection{Preliminary consistency conditions}
\label{sec:preconsistency}

In principle, one might consider the basic fields $\{g_{\mu\nu},T_\mu,S_\mu\}$, when promoted to distributions, to have an arbitrary regular part and some singular parts. Regarding the two vectors, the equations of motion are plagued by products involving several $T_\mu$ and/or $S_\mu$. Therefore, it is reasonable to assume from the beginning that they are regular tensor distributions (see Definition \ref{def:regularten}), namely,
\begin{align}
    \udis{T}_\mu &= T^{+}_\mu \udis{\Theta}^{+} + T^{-}_\mu \udis{\Theta}^{-}\,,\\
    \udis{S}_\mu &= S^{+}_\mu \udis{\Theta}^{+} + S^{-}_\mu \udis{\Theta}^{-}\,,
\end{align}
so as to avoid both products of Dirac deltas and higher-order singular contributions. For the case of the metric, as explained previously, we assume it to be a regular and continuous tensor across $\Sigma$.

\subsubsection{Ill-defined combinations from torsion equations}

In this subsection we shall prove that no consistency condition is obtained from the torsion equations \eqref{eq:EoM1} and \eqref{eq:EoM2}. To see it, let us represent by $V_\mu$ any of the variables $T_\mu$ or $S_\mu$. All the terms of the gravitational part of the equations \eqref{eq:EoM1} and \eqref{eq:EoM2} have one of the following schematic structures (we drop coefficients as well as the indices of the vectors $T_\mu$ and $S_\mu$ and their derivatives):
\begin{align}
    \LCD R &= \LCD \LCR + \LCD\LCD V + V \LCD V  \,, \\
    \LCD \mathcal{H} &= \LCD\LCD V + V\LCD V \,,\\
    V\mathcal{H} &=  V\LCD V + VVV\,,\\
    VR &= V\LCR + V\LCD V + VVV\,,\\
    V\,.\!\!\phantom{R}&
\end{align}
We clearly see that, when promoted to distributions, none of the resulting terms above involve quadratic terms in singular parts. Therefore, we conclude that the torsion equations \eqref{eq:EoM1} and \eqref{eq:EoM2} can be regarded as well-defined in the distributional sense under the preliminary conditions stated in Section \ref{sec:preconsistency} and the prescription \eqref{eq:prescTheDel}.

\subsubsection{Ill-defined combinations from the equation of the metric}

As done for the equations \eqref{eq:EoM1} and \eqref{eq:EoM2}, we now present the schematic form of all the terms in the gravitational part of the metric equation \eqref{eq:EoM3} (where, again, irrelevant indices and coefficients are dropped out):
\begin{align}
    \LCR_{\mu\nu} R &= \LCR_{\mu\nu} \LCR + \LCR_{\mu\nu}\LCD V + \LCR_{\mu\nu}VV\,, \label{eq:problems1}\\
    \LCD\LCD R &= \LCD\LCD \LCR + \LCD\LCD\LCD V + \LCD V \LCD V +  V \LCD\LCD V \,, \label{eq:problems2}\\
    R^2 &= (\LCR + \LCD V + V V)^2 = \LCR\LCR + \LCR\LCD V +\LCD V \LCD V+ \LCR VV + V V\LCD V +VVVV  \,,\\
    \mathcal{H}^2&=(\LCD V + V V)^2 =  \LCD V \LCD V+ \LCR VV + V V\LCD V +VVVV \,,\\
    V\LCD R &= V\LCD\LCD \LCR + V\LCD\LCD V + VV \LCD V  \,,\\
    V\LCD \mathcal{H} &= V\LCD\LCD V + VV\LCD V \,,\\
    VV\mathcal{H} &=  VV\LCD V + VVVV\,,\\
    VVR &= VV\LCR + VV\LCD V + VVVV\,,\\
    VV\,, & \\
    \LCR_{\mu\nu}\,.&
\end{align}
Notice that only the first four expressions above would give quadratic terms in the Dirac delta. In particular, the problematic combinations are $\LCR_{\mu\nu} \LCR$, $\LCR \LCR$, $\LCD V\LCD V$, $\LCR_{\mu\nu} \LCD V$ and $\LCR \LCD V$. Thus, let us analyze them in full detail to find the corresponding consistency conditions.

We first focus on the terms $\beta R^2$ and $\alpha \mathcal{H}^2$ in the equation   \eqref{eq:EoM3}. After expanding them in terms of $\{g_{\mu\nu},T_\mu,S_\mu\}$, we find:
\begin{align}
    \beta R^2 &= \beta \LCR^2+4\beta (\LCD_\mu T^{\mu})^2+\ldots\\
    \alpha\mathcal{H}^2 &= \alpha (\LCD_\mu S^{\mu})^2+\ldots
\end{align}
To be well-defined in the sense of distributions, neither $\udis{\LCR}$, $\disLCD_\mu \udis{T}^{\mu}$ nor $\disLCD_\mu \udis{S}^{\mu}$ can have a non-trivial singular part. Due to the conditions in Section \ref{sec:preconsistency}, the singular part of these three distributions is of the lowest order, i.e., proportional to the Dirac delta. The nullity of these pieces leads us, respectively, to impose the following conditions
\begin{align}
        \beta\jump{K}=0, \label{eq:continuity of beta K} \\
        \beta\jump{T_\perp}=0, \\
        \alpha\jump{S_\perp}=0. \label{eq:continuity of alpha S}
\end{align}
Notice that we included here the parameters to avoid branching our analysis in different subcases depending on whether $\alpha$ and/or $\beta$ vanish or not (we will comment on these particular scenarios in Section \ref{sec:partcases}). 

Let us now focus on the two remaining problematic structures, namely \eqref{eq:problems1} and \eqref{eq:problems2}. Owing to \eqref{eq:continuity of beta K}-\eqref{eq:continuity of alpha S}, the term \eqref{eq:problems1} is well-defined in the sense of distributions, so we are left with \eqref{eq:problems2}. In this term, only the combinations $\LCD V\LCD V$ are potentially problematic, so let us compute them:
\begin{equation}
    \beta \LCD_\mu \LCD_\nu R = \frac{1}{12}\beta \LCD_\mu S^\rho \LCD_\nu S_\rho -\frac{4}{3}\beta \LCD_\mu T^\rho \LCD_\nu T_\rho + \ldots
\end{equation}
An analysis of the previous expression reveals that, for consistency, we have to impose two further conditions, namely:
\begin{align}
        \beta\jump{T_\mu}=0\,, \\
        \beta\jump{S_\mu}=0\,. 
\end{align}

\subsubsection{Summary of consistency conditions}
\label{cons:cond}

As per the results obtained thus far in the previous subsections, the conditions required so as to avoid ill-defined terms in the distributional equations of motion are as follows:
\begin{align}
    0&=\singpart{\ell}[T_\mu]=\singpart{\ell}[S_\mu]=\singpart{\ell}[g_{\mu\nu}]\mybigskip \forall \ell=0,1,...\,; \label{eq:condNoSingTSg}\\
    0&=\jump{g_{\mu\nu}}\,; \label{eq:condnojumpg}\\
    0&=\beta\jump{K}
    \,\,\mybigskip\left(\Leftrightarrow\quad \singpart{0}[\beta\udis{\LCR}]=0 \right)\,; \label{eq:condJumpK0}\\
    0&=\beta\jump{T_\mu}
    \,\mybigskip\left(\Leftrightarrow\quad \singpart{0}[\beta\disLCD_\mu \udis{T}_\nu]=0 \right)\,; \label{eq:condbjumpT0}\\
    0&=\beta\jump{S_\mu}
    \,\mybigskip\left(\Leftrightarrow\quad \singpart{0}[\beta\disLCD_\mu \udis{S}_\nu]=0 \right)\,; \label{eq:condbjumpS0}\\
    0&=\alpha\jump{S_\perp}
    \mybigskip\left(\Leftrightarrow\quad \singpart{0}[\alpha\disLCD_\mu \udis{S}^\mu]=0 \right)\,. \label{eq:condJumpnS}
\end{align}

\subsection{Decomposition of the equations of motion}
\label{sec:decomp}

In this subsection, we explore the singular pieces of both sides of the equations of motion \eqref{eq:EoM1}-\eqref{eq:EoM3} in order for them to be consistent.

First, we have to analyze carefully each of the terms in all of the equations, so as to see the type of singular contributions that every equation may contain \textit{after} imposing consistency conditions \eqref{eq:condNoSingTSg}-\eqref{eq:condJumpnS}. The results for each of the equations are tabulated in Appendix \ref{app:tablesterms}. By inspection, we can deduce that the matter sides of \eqref{eq:EoM1}-\eqref{eq:EoM3} are only allowed to have the following non-vanishing pieces
\begin{align}
    \udis{\trcurrent}_\mu  &= \regpart\left[\udis{\trcurrent}_\mu\right]+\singpart{0}[\udis{\trcurrent}_\mu] \,, \\
    \udis{\axcurrent}_\mu  &= \regpart\left[\udis{\axcurrent}_\mu\right]\, +\singpart{0}[\udis{\axcurrent}_\mu]\,, \\
    \udis{T}_{\mu\nu} &= \regpart\left[\udis{T}_{\mu\nu}\right]+\singpart{0}[\udis{T}_{\mu\nu}] +\singpart{1}[\udis{T}_{\mu\nu}]\,.
\end{align}
Let us now study each of them separately.

\subsubsection{Singular parts of the vector spin density}
\label{sing:trace}

We start with the singular contributions to the equation of the torsion trace vector $T_\mu$, which will determine $\singpart{0}[\udis{\trcurrent}_\mu]$. Keeping in mind the results in Table \ref{tab:termsEqT}, and rearranging the gravitational part, equation \eqref{eq:EoM1} can be expressed as
\begin{align}    
    \trcurrent_\mu & = -4\beta\LCD_\mu R + \text{(regular tensor terms)} \nonumber\\
          & = -4\beta\LCD_\mu \LCR - 8\beta \LCD_\mu \LCD_\nu T^\nu + \text{(regular tensor terms)}\,.
\end{align}
Now that all terms in the previous expression are given in terms of the basic fields and their derivatives, we can promote them to distributions:
\begin{equation}    
    \udis{\trcurrent}_\mu = -4\beta\disLCD_\mu \udis{\LCR} - 8\beta \disLCD_\mu \disLCD_\nu \udis{T}^\nu + \text{(regular distributional terms)}\,.
\end{equation}
Hence,
\begin{equation}    
    \singpart{0}[\udis{\trcurrent}_\mu]  = -4\beta\singpart{0}[\disLCD_\mu \udis{\LCR}] - 8\beta \singpart{0}[\disLCD_\mu \disLCD_\nu \udis{T}^\nu]\,.
\end{equation}
Finally, we use the traces of \eqref{eq:Delta0DDV} and \eqref{eq:jumpDV}, as well as \eqref{eq:Delta0DR} and
\begin{equation}
    \beta\jump{\LCR} +2\epsilon\beta n^\mu n^\nu \jump{\LCD_\mu T_\nu}=\beta\jump{R}\label{eq:jumpRsimp}
\end{equation}
(of course, keeping in mind \eqref{eq:condJumpK0} and \eqref{eq:condbjumpT0}), to get:
\begin{equation} 
\label{D0Lmu}
    \singpart{0}[\udis{\trcurrent}_\mu]  = -4\epsilon \beta \jump{R} n_\mu \del \,.
\end{equation}
In conclusion, configurations without monopole density (thin shell) in $\trcurrent_\mu$  are allowed for $\beta\neq0$ if and only if $\jump{R}=0$.

\subsubsection{Singular parts of the axial vector spin density}
\label{sing:axial}

We now turn to study the singular contribution to the equation of the torsion axial vector, which will determine $\singpart{0}[\udis{\axcurrent}_\mu]$, following the same approach as in the previous subsection. The information provided in Table \ref{tab:termsEqS} 
allows us to expand $J_{\mu}$ as
\begin{align}    
    \axcurrent_\mu & = 2\alpha\LCD_\mu \mathcal{H} + \text{(regular tensor terms)} \nonumber\\
          & = -2\alpha \LCD_\mu \LCD_\nu S^\nu +\frac{4\alpha}{3}(T^\nu \LCD_\mu S_\nu + S^\nu \LCD_\mu T_\nu) + \text{(regular tensor terms)}\,.
\end{align}
We now promote to distributions, obtaining
\begin{equation}    
    \udis{\axcurrent}_\mu = -2\alpha \disLCD_\mu \disLCD_\nu \udis{S}^\nu +\frac{4\alpha}{3}(\udis{T}^\nu \disLCD_\mu \udis{S}_\nu + \udis{S}^\nu \disLCD_\mu \udis{T}_\nu) + \text{(regular distributional terms)}\, ,
\end{equation}
where the prescription \eqref{eq:prescTheDel} has been assumed. Therefore, we have
\begin{equation}    
    \singpart{0}[\udis{\axcurrent}_\mu]  = -2\alpha \singpart{0}[\disLCD_\mu \disLCD_\nu \udis{S}^\nu] +\frac{4\alpha}{3}\left(T^{\Sigma\nu} \singpart{0}[\disLCD_\mu \udis{S}_\nu] + S^{\Sigma\nu} \singpart{0}[\disLCD_\mu \udis{T}_\nu]\right)\,.
\end{equation}

By using the traces of \eqref{eq:Delta0DDV} and \eqref{eq:jumpDV}, as well as \eqref{eq:decompDV} (keeping in mind \eqref{eq:condJumpnS}), we get:
\begin{equation}    
    \singpart{0}[\udis{\axcurrent}_\mu]  =    2\epsilon \alpha \jump{\mathcal{H}} n_\mu  \del\,.
\end{equation}
We can then conclude that configurations without monopole density in $\axcurrent_\mu$ would require $\jump{\mathcal{H}}=0$ provided $\alpha\neq 0$.

\subsubsection{Singular parts of the energy-momentum tensor}

\label{sing:Einstein}

Finally, we examine the singular pieces of the energy-momentum tensor. The information in Table \ref{tab:termsEqMetric} leads us to
\begin{align}    
    T_{\mu\nu} & = \Mp^2 \LCR_{\mu\nu} + 4 \beta \LCR_{\mu\nu} \LCR + \frac{1}{6} \beta \LCR_{\mu\nu} S_\sigma S^\sigma -  \frac{8}{3} \beta \LCR_{\mu\nu} T_\sigma T^\sigma \nonumber\\
    &\quad
    + 8 \beta \LCR_{\mu\nu} \LCD_\sigma T^\sigma - 4 \beta \LCD_\mu\LCD_\nu \LCR - 8 \beta \LCD_\mu\LCD_\nu\LCD_\sigma T^\sigma \nonumber\\
    &\quad - \frac{1}{3} \beta S^\sigma \LCD_{(\mu}\LCD_{\nu)} S_\sigma + \frac{16}{3} \beta T^\sigma \LCD_{(\mu}\LCD_{\nu)} T_\sigma - 8 \beta T_{(\mu} \LCD_{\nu)} \LCR - 16 \beta T_{(\mu} \LCD_{\nu)}\LCD_\sigma T^\sigma \nonumber\\
    &\quad+ \frac{8}{3} \alpha S^\sigma S_{(\mu} \LCD_{\nu)} T_\sigma+ \frac{8}{3} \alpha T^\sigma S_{(\mu} \LCD_{\nu)} S_\sigma -  4\alpha S_{(\mu} \LCD_{\nu)}\LCD_\sigma S^\sigma  \nonumber\\   
    & \quad + 2 g_{\mu\nu} \Big(-\frac{\Mp^2}{4} \LCR + 2 \beta T^\sigma \LCD_\sigma\LCR  + 2 \beta  \LCbox \LCR + 4 \beta T^\sigma \LCD_\sigma\LCD_{\lambda}T^{\lambda}\nonumber\\
    &\qquad\qquad+ \alpha S^\sigma \LCD_\sigma\LCD_{\lambda}S^{\lambda} -  \frac{2}{3} \alpha S^\sigma T^{\lambda} \LCD_\sigma S_{\lambda} -  \frac{2}{3} \alpha S^\sigma S^{\lambda} \LCD_{\lambda}T_\sigma \nonumber\\
    &\qquad\qquad+ \frac{1}{6} \beta S^\sigma \LCbox S_\sigma -  \frac{8}{3} \beta T^\sigma \LCbox T_\sigma + 4 \beta  \LCbox \LCD_\sigma T^{\sigma}\Big)
+ \text{(regular tensor terms)}\, . \label{eq:Tmnsing}
\end{align}

In this case, the singular part consists of a dipolar (double-layer) and a monopolar (thin-shell) component, which we shall analyze separately in the following.

\paragraph*{\textbf{Dipole.}}
When promoted to distributions, only the following terms in the stress-energy tensor provide double-layer contributions:
\begin{equation}    
    \singpart{1}[\udis{T}_{\mu\nu}] = - 4 \beta (\delta^\alpha_\mu \delta^\beta_\nu - g_{\mu\nu} g^{\alpha\beta}) \left(
    \singpart{1}[\disLCD_\alpha \disLCD_\beta \udis{\LCR}] + 2 \singpart{1}[\disLCD_\alpha\disLCD_\beta \disLCD_\sigma \udis{T}^\sigma]\right)\,,
\end{equation}
which, according to \eqref{eq:decomDDR2} and \eqref{eq:decomDDDT2}, and making use of \eqref{eq:jumpRsimp}, gives
\begin{equation}    
    \singpart{1}[\udis{T}_{\mu\nu}]  =  4\epsilon \beta\, \disLCD_\sigma \! \left(
     \jump{R} h_{\mu\nu} n^\sigma \del \right)\,.
\end{equation}

\paragraph*{\textbf{Monopole.}}
All the explicit terms in \eqref{eq:Tmnsing} will contribute to the monopolar part $\singpart{0}[\udis{T}_{\mu\nu}]$. We split it into three parts:
\begin{equation}
    \singpart{0}[\udis{T}_{\mu\nu}] \equiv \epsilon\left(\Mp^2\, H^{(\text{GR})}_{\mu\nu} + \beta\,  H^{(\beta)}_{\mu\nu} + \alpha\, H^{(\alpha)}_{\mu\nu}\right) \del \,,\label{eq:sin0THs}
\end{equation}
with the different contributions being
\begin{align}
    H^{(\text{GR})}_{\mu\nu} &=  h_{\mu\nu}\jump{K} - \jump{K_{\mu\nu}}\,,\label{eq:HGR}\\
    H^{(\beta)}_{\mu\nu} &=  -4 \jump{K_{\mu\nu}}R^\sSigma -4 \Big(K^\sSigma_{\mu\nu} -\epsilon n_\mu n_\nu K^\sSigma+2 n_{(\mu}T^\sSigma_{\nu)}- g_{\mu\nu} T^\sSigma_{\perp} \Big) \jump{R} \nonumber\\
    &\quad -4 (2 n_{(\mu}h^\rho{}_{\nu)}- h_{\mu\nu}n^\rho)\partial_\rho \jump{R} \,,\label{eq:Hbeta}\\
    H^{(\alpha)}_{\mu\nu} &= 2\left(2 n_{(\mu} S^\sSigma_{\nu)} - g_{\mu\nu}  S^\sSigma_\perp\right) \jump{\mathcal{H}}\,.\label{eq:Halpha}
\end{align}
Here, one should bear in mind that\footnote{
        Observe that, in the expression for $R^\sSigma$, one may use \eqref{eq:ssigmaproduct} to simplify $\beta(S_\rho S^{\rho})^\sSigma=\beta S^\sSigma_\rho S^{\sSigma\rho}$ and $\beta(T_\rho T^{\rho})^\sSigma=\beta T^\sSigma_\rho T^{\sSigma\rho}$, because both  $S_\mu$ and $T_\mu$ are continuous when multiplied by $\beta$. }
\begin{equation}
    R^\sSigma = \LCR^\sSigma + 2 (\LCD_\rho T^{\rho})^\sSigma + \frac{1}{24}(S_\rho S^{\rho})^\sSigma - \frac{2}{3}(T_\rho T^{ \rho})^\sSigma.
\end{equation}
Moreover, in deriving \eqref{eq:Hbeta}, we made use of the following useful simplification (where $V$ represents either $T$ or $S$):
\begin{equation}
    \beta\jump{\LCD_\rho V^\lambda}V^\sSigma_\lambda = \frac{\beta}{2}\jump{\LCD_\rho (V^\lambda V_\lambda)} + \underbrace{(...)\beta\jump{V_\mu}}_0 = \frac{1}{2}\partial_\rho\jump{\beta(V^\lambda V_\lambda)} = \partial_\rho(\underbrace{\beta\jump{V^\lambda}}_0 V^\sSigma_\lambda) = 0\,.
\end{equation}

\subsubsection{Summary of results}

Now we present a summary of the singular parts of the equations of motion. If we use the notation in \eqref{eq:defD0}-\eqref{eq:defD1} to write
\begin{align}
    \singpart{0}[\udis{\trcurrent}_\mu] &=\trcurrent^\szero_\mu\del\,,
    & \singpart{0}[\udis{T}_{\mu\nu}] &=T^\szero_{\mu\nu}\del\,,\nonumber\\   
    \singpart{0}[\udis{\axcurrent}_\mu] &=\axcurrent^\szero_\mu\del\,,
    &\singpart{1}[\udis{T}_{\mu\nu}]&=\disLCD_\rho (T^\sone_{\mu\nu} n^\rho \del)\,,
\end{align}
then the non-vanishing multipole densities in $\Sigma$ are
\begin{align}
    \trcurrent^\szero_\mu &= -4\epsilon \beta \jump{R} n_\mu\,, \\
    \axcurrent^\szero_\mu &=    2\epsilon \alpha \jump{\mathcal{H}} n_\mu \,,\\
    T^\szero_{\mu\nu} & =\epsilon\left(\Mp^2\, H^{(\text{GR})}_{\mu\nu} + \beta\,  H^{(\beta)}_{\mu\nu} + \alpha\, H^{(\alpha)}_{\mu\nu}\right)\,,\\
    T^\sone_{\mu\nu} & =  4\epsilon \beta
     \jump{R} h_{\mu\nu} \,, \label{eq:stress-energy double layer}
\end{align}
where $H^{(\text{GR})}_{\mu\nu}$, $H^{(\beta)}_{\mu\nu}$ and $H^{(\alpha)}_{\mu\nu}$ are respectively given in \eqref{eq:HGR}, \eqref{eq:Hbeta} and \eqref{eq:Halpha}. Observe that, if the monopole density of $\trcurrent_\mu$ is zero, the dipole density of $T_{\mu\nu}$ vanishes identically.

It is also convenient to decompose $T^\szero_{\mu\nu}$ into its normal and tangential components,
\begin{equation}
    T^\szero_{\mu\nu} = \tau n_\mu n_\nu + 2 n_{(\mu}\tau_{\nu)}+\tau_{\mu\nu}\,,
\end{equation}
where we are following the notation in \cite{Reina:2015gxa}:
\begin{equation}
    \tau\equiv n^\mu n^\nu T^\szero_{\mu\nu}\,,\mybigskip
    \tau_\mu\equiv \epsilon h^\lambda{}_\mu n^\nu T^\szero_{\mu\nu}\,,\mybigskip
    \tau_{\mu\nu} \equiv h^\sigma{}_\mu h^\rho{}_\nu T^\szero_{\sigma\rho}\,.
\end{equation}
Notice that, by definition, the last two objects are purely tangential, i.e.~$n^\mu\tau_\mu=0=n^\mu\tau_{\mu\nu}$. The splitting of the monopolar part of the energy-momentum tensor then reads:
\begin{align}
    \tau &= 2\alpha S_\perp^\sSigma \jump{\mathcal{H}} - 4 \beta (T_\perp^\sSigma-K^\sSigma) \jump{R}\,,\nonumber\\
    \tau_\mu &=  2\epsilon\alpha\tosigma{S}^\sSigma_\mu \jump{\mathcal{H}}  -4\epsilon \beta \left(\tosigma{T}^\sSigma_\mu \jump{R} + h^\sigma{}_\mu \partial_\sigma\jump{R}\right)\,, \nonumber\\
    \tau_{\mu\nu} &= \Mp^2\epsilon \big( h_{\mu\nu}\jump{K} - \jump{K_{\mu\nu}}\big) -2\epsilon\alpha h_{\mu\nu} S^\sSigma_\perp \jump{\mathcal{H}}  \nonumber\\
    & \quad -4\epsilon\beta \Big[\jump{K_{\mu\nu}}R^\sSigma +\Big(K^\sSigma_{\mu\nu} - h_{\mu\nu} T^\sSigma_{\perp} \Big)\jump{R}  - h_{\mu\nu}n^\rho\partial_\rho \jump{R}\Big]\,.\label{eq:Tdecomp}
\end{align}

\subsection{Particular cases}
\label{sec:partcases}
Having obtained the generic junction conditions for arbitrary $\alpha$ and $\beta$, we will now focus on some paradigmatic subcases in which these parameters take specific values, and compare the results with the existing literature.

\subsubsection{Theories without the Ricci-square term (\texorpdfstring{$\beta=0$}{β=0})}
\label{sec:betaeq0}

An immediate consequence of expression \eqref{eq:stress-energy double layer} is that all PG models with $\beta=0$ are incompatible with double-layer configurations, because the dipolar part of the energy-momentum tensor is always proportional to $\beta(=0)$. 

There are two non-trivial cases with $\beta =0$. First, let us assume that, in addition to vanishing $\beta$, we also have $\alpha=0$. In such a scenario, we end up with the Einstein-Cartan theory with mass terms for the torsion pieces. There are no extra consistency conditions besides the regularity of the vectors and the continuity of the metric. Also, both vector currents become purely regular (i.e., there are no surface contributions to them), and only the energy-momentum tensor has a singular part, which coincides with the one it has in GR:
\begin{equation}
    T^\szero_{\mu\nu}  = \epsilon \Mp^2 (h_{\mu\nu}\jump{K} - \jump{K_{\mu\nu}})\,.
\end{equation}

If we now focus on the other possibility, namely $\beta=0$ but $\alpha\neq 0$, we get one extra consistency condition:
\begin{equation}
\label{jumpSperp}
    \jump{S_\perp}=0.
\end{equation}
However, the condition \eqref{jumpSperp} does not help to further simplify any expression. The only non-trivial singular pieces are:
\begin{align}
    \axcurrent^\szero_\mu &=   2\epsilon\alpha \jump{\mathcal{H}} n_\mu  \,, \nonumber\\
    T^\szero_{\mu\nu}  &= \epsilon\left(\Mp^2\, H^{(\text{GR})}_{\mu\nu} + \alpha\,  H^{(\alpha)}_{\mu\nu}\right)  \,.
\end{align}
Thus, we find that the three pieces $\tau$, $\tau_\mu$ and $\tau_{\mu\nu}$, as defined in \eqref{eq:Tdecomp}, are, in general, non-trivial. Consequently, the absence of surface spin densities and matter thin shells at $\Sigma$ ($\axcurrent^\szero_\mu = \tau= \tau_\mu = \tau_{\mu\nu}=0$) leads to the following smooth-matching conditions:
\begin{equation}
    \jump{S_\perp}=0\,, \mybigskip \jump{\mathcal{H}}=0\,,\mybigskip\jump{K_{\mu\nu}}=0\,.\label{eq:JCbeq0aneq0}
\end{equation}

\subsubsection{Theories with Ricci-square term (\texorpdfstring{$\beta\neq0$}{β≠0})}
\label{sec:betaneq0}

For $\beta\neq 0$, besides the continuity of the metric and the regularity of the torsion vectors, we also obtain:
\begin{equation}
    \jump{S_\mu}=\jump{T_\mu}=\jump{K}=0, \label{eq:condbneq0}
\end{equation}
i.e., the three basic fields of the theory $\{g_{\mu\nu}, T_\mu,S_\mu\}$, as well as the trace of the extrinsic curvature $K$, are all continuous. If, on top of that, $\alpha\neq 0$, we get an additional condition ($\jump{S_\perp}=0$), but the latter is redundant, so \eqref{eq:condbneq0} are in fact valid for arbitrary $\alpha$.

Consequently, for non-vanishing $\beta$ and arbitrary $\alpha$, the only non-trivial singular pieces of the equations of motion become:
\begin{align}
    \trcurrent^\szero_\mu&= -4\epsilon\beta \jump{R} n_\mu \,, \nonumber\\
    \axcurrent^\szero_\mu &= 2\epsilon\alpha \jump{\mathcal{H}} n_\mu \,, \nonumber\\
    T^\szero_{\mu\nu}  &= -\epsilon\Mp^2\jump{K_{\mu\nu}} + \epsilon\beta\,  H^{(\beta)}_{\mu\nu} + 2\epsilon\alpha\left(2 n_{(\mu} S^\sSigma_{\nu)} - g_{\mu\nu}  S^\sSigma_\perp\right) \jump{\mathcal{H}} \,, \nonumber\\
    T^\sone_{\mu\nu} & =  4\epsilon \beta
     \jump{R} h_{\mu\nu} \,. \label{eq:JCbneq0}
\end{align}
Notice that, under the consistency conditions \eqref{eq:condbneq0}, we find:
\begin{equation}
    \jump{\mathcal{H}} = -\epsilon n^\nu n^\rho \jump{\LCD_\nu S_\rho}\,.
\end{equation}

We observe that a non-vanishing monopolar contribution to $\trcurrent_\mu$ leads to a non-trivial double layer in the energy-momentum tensor. Moreover, the monopolar parts of both $\trcurrent_\mu$ and $\axcurrent_\mu$ contribute to the energy-momentum thin shell. Actually, even if they are both vanishing, i.e., if $\jump{R} =\alpha\jump{\mathcal{H}}=0$, the torsion vectors continue to have an impact in the energy-momentum monopole (in particular, in the tangential energy-momentum, via $R^\sSigma$):
\begin{equation}
    \tau = 0\,, \mybigskip
    \tau_\mu = 0 \,, \mybigskip
    \tau_{\mu\nu} = -\epsilon (\Mp^2+4\beta R^\sSigma)\jump{K_{\mu\nu}}\,.
\end{equation}
Therefore, the absence of spin density monopoles and thin shells is guaranteed if and only if
\begin{equation}
    \jump{R} = \alpha\jump{\mathcal{H}}=0\,,\mybigskip  (\Mp^2+4\beta R^\sSigma)\jump{K_{\mu\nu}}=0\,.
\end{equation}

Bi-scalar Poincaré Gauge gravity corresponds to the case $\alpha\neq 0$, which leads to continuous total Ricci scalar and Holst pseudo-scalar. In addition, we get the generic condition $\jump{K_{\mu\nu}}=0$, except for those configurations for which $\Mp^2+4\beta R^\sSigma=0$.

In principle, we could compare our results for $\alpha=0$ (which corresponds to the Palatini $R+R^2$ limit without nonmetricity) with the ones obtained in \cite{Olmo:2020fri,Rosa:2021mln}. Nevertheless, the formalism employed in both references requires some manipulations of the various quantities which is not compatible with our approach\footnote{
    Throughout this article, we have presented some ambiguities and subtleties that appear when studying the junction conditions of a theory using distributions. This implies that each author would need to make certain choices to deal with them, implying that different formalisms might not be compatible with each other, and hence the results cannot be compared.}
as described in Section \ref{sec:procedure}. In particular, \cite{Olmo:2020fri,Rosa:2021mln} assume vanishing hypermomentum currents and integrate the equation of motion of the connection, resulting in denominators containing quantities that are to be promoted to distributions.

\subsubsection{Metric Ricci-square gravity theories}

After a careful look, one may immediately realize that the metric Ricci-square theory can be recovered from the results in Section \ref{sec:betaneq0} by dropping the torsion (as well as its associated equations and masses). The only consistency condition in this case, in addition to the continuity of the metric, is
\begin{equation}
    \jump{K}=0.
\end{equation}
Furthermore, the only matter current in the metric-Ricci-square case, the energy-momentum, has the following singular pieces:
\begin{align}
    T^\szero_{\mu\nu} &= -\epsilon\bigg[\Mp^2\,\jump{K_{\mu\nu}} +4 \beta\bigg( \jump{K_{\mu\nu}}\LCR^\sSigma + \jump{\LCR}\, (K^\sSigma_{\mu\nu} -\epsilon n_\mu n_\nu K^\sSigma) \nonumber\\
    &\qquad\qquad +(2n_{(\mu}h^\rho{}_{\nu)}- h_{\mu\nu}n^\rho) \partial_\rho \jump{\LCR} \bigg)     
    \bigg]\,, \nonumber\\
T^\sone_{\mu\nu} & =  4\epsilon \beta
     \jump{R} h_{\mu\nu}\,.
\end{align}

Hence, in order to compare with the results of metric Ricci-square gravity in \cite{Senovilla:2013vra}, we will take into account the decomposition of the monopolar part given in \eqref{eq:Tdecomp}, since the results of the aforementioned reference are presented in that form. Consequently:
\begin{align}
    \tau &= 4\beta K^\sSigma \jump{\LCR}\,,\nonumber\\
    \tau_\mu &= -4\beta\epsilon h^\sigma{}_\mu \partial_\sigma\jump{\LCR}\,, \nonumber\\
    \tau_{\mu\nu} &= -\epsilon (\Mp^2+4\beta \LCR^\sSigma)\jump{K_{\mu\nu}} -4\epsilon\beta \Big(K^\sSigma_{\mu\nu} \jump{\LCR} - h_{\mu\nu}n^\rho\partial_\rho \jump{\LCR}\Big)\,.
\end{align}
These results agree with the ones in the Appendix of \cite{Senovilla:2013vra}, after adapting the conventions: $\epsilon\to 1$, $\Mp^2\to 1/\kappa $, $\beta \to \alpha/(2\kappa)$ and $n^\rho \partial_\rho \jump{\LCR}\to a$.

\subsection{Comments on matter sectors depending on the Vielbein}
\label{sec:Dirac}

Interestingly, the paradigmatic matter Lagrangian with non-trivial spin density, the Dirac Lagrangian, does not (automatically at least) fit in the formalism developed in previous sections. The reason lies in the fact that such a matter Lagrangian requires additional geometrical structures, forcing one to work in the Vielbein formulation. Accordingly, the metric should not be a basic field of the theory but the Vielbein $e^I{}_\mu$ (or equivalently, the inverse Vielbein $e_I{}^\mu$), which satisfy:
\begin{equation}
    g_{\mu\nu} = e^I{}_\mu e^J{}_\nu\, \eta_{IJ}\,,\mybigskip g^{\mu\nu} = e_I{}^\mu e_J{}^\nu\, \eta^{IJ} \,,
\end{equation}
where $\eta$ denotes the Minkowski metric in Cartesian coordinates. At the level of the action, we have:
\begin{equation}
    \tilde{\mathcal{S}}[e^I{}_\mu, T_\mu, S_\mu, t^\rho{}_{\mu\nu}, \Psi]=\mathcal{S}_\text{g}[g^{\mu\nu}(e^I{}_\mu), T_\mu, S_\mu, t^\rho{}_{\mu\nu}] + \mathcal{S}_\text{m}[e^I{}_\mu, T_\mu, S_\mu, \Psi]\,.
\end{equation}
Now, instead of the equation of motion for the metric, we get:
\begin{equation}
    0=\frac{1}{|e|}\frac{\delta \tilde{\mathcal{S}}}{\delta e^I{}_\mu} = \frac{1}{|e|}\frac{\delta \mathcal{S}_\text{g}}{\delta g^{\alpha\beta}}\left(-g^{\mu (\beta} 2 e_I{}^{\alpha)}\right) + \tilde{T}^\mu{}_{I}
    = \left(\mathcal{E}(g)_{\alpha\beta}-\tilde{T}_{\alpha \beta}\right)\left(-g^{\mu \alpha} e_I{}^{\beta}\right), \label{eq:EoMvielbein}
\end{equation}
where $\mathcal{E}(g)_{\alpha\beta}$ correspond to the standard metric variation (see \eqref{eq:var0}), $e \equiv \det (e^I{}_\mu)$ and
\begin{equation}
    \tilde{T}^\mu{}_{I}\equiv \frac{1}{|e|}\frac{\delta \tilde{\mathcal{S}}}{\delta e^I{}_\mu}\,,\mybigskip \tilde{T}_{\alpha \beta} \equiv \tilde{T}^\mu{}_{I}\, g_{\mu\alpha}\, e^I{}_\beta\,.
\end{equation}
After contracting with the Vielbein and the metric, we can split the equation \eqref{eq:EoMvielbein} into symmetric and antisymmetric parts:
\begin{equation}
    \mathcal{E}(g)_{\mu\nu}=\tilde{T}_{(\mu\nu)}\,,\mybigskip \tilde{T}_{[\mu\nu]}=0\,.
\end{equation}
In the first equation, we can straightforwardly use all the previous results for symmetric energy-momentum tensors, since the gravitational part $\mathcal{E}(g)_{\mu\nu}$ is the same as in metric formulation. The second equation, however, just tells us that all the singular parts coming from $\tilde{T}_{[\mu\nu]}$ must vanish identically.

\section{Conclusions}
\label{sec:conclusions}

In this work we have obtained the junction conditions for a ghost-free subclass of PG gravity which propagates one scalar and one pseudo-scalar. We have given new insights on the formalism of junction conditions in gravitational theories in Section \ref{sec:procedure}. Then, we have particularized our study to the mentioned theory. Since we have not assumed any specific value for the parameters of the theory, the results are completely general. 

The junction conditions have been obtained in two steps. First, in order to avoid products of singular distributions, we have determined the consistency conditions of the theory, which can be found in Section \ref{cons:cond}. Then, we have derived the relation between the matter content of the hypersurface and the allowed discontinuities in the fields of the theory under the previously obtained consistency conditions. For such a purpose, we have first studied the singular contributions to the equations of motion, which are collected in Tables \ref{tab:termsEqT}-\ref{tab:termsEqMetric}. Then, in  Section \ref{sing:trace} we have analyzed the case of the trace vector equation and showed that the only singular part allowed for the corresponding matter current is a thin shell controlled by the combination $\beta\jump{R}$. In Section \ref{sing:axial}, we have carried out an analogous analysis for the axial vector equation, finding once again that only thin shells are allowed, with the corresponding singular part being characterized by the product $\alpha\jump{\mathcal{H}}$. Then, concerning the equation of the metric (Section \ref{sing:Einstein}), we have shown that both thin-shell (monopolar) and double-layer (dipolar) singular contributions are allowed for the energy-momentum tensor. The dipolar part of the energy-momentum tensor vanishes identically for $\beta\jump{R}=0$. On the other hand, the corresponding thin shell has non-vanishing irreducible parts (tension scalar $\tau$, external momentum flux $\tau_\mu$ and surface energy-momentum tensor $\tau_{\mu \nu}$). The first two contain terms proportional to $\beta\jump{R}$ and $\alpha\jump{\mathcal{H}}$, while $\tau_{\mu \nu}$ acquires a contribution proportional to $\beta \jump{K_{\mu\nu}}$, in addition to its GR part and terms proportional to $\beta\jump{R}$ and $\alpha\jump{\mathcal{H}}$.

Finally, we have studied different sub-cases of the theory in Section \ref{sec:partcases}:
\begin{itemize}
    \item For $\alpha=\beta=0$, we have found the same junction conditions as in GR. For $\beta=0$ but $\alpha\neq0$,  whose junction conditions are collected in \eqref{eq:JCbeq0aneq0}, we have shown that the absence of thin shells of any kind require the continuity of $S_\perp$, $\mathcal{H}$ and $K_{\mu\nu}$. Moreover, condition $\beta=0$ implies that no double layers are possible in these models.

    \item For $\beta\neq 0$ and arbitrary $\alpha$, we have proved that we have to impose the continuity of $S_\mu$, $T_\mu$ and $K$ in order to avoid ill-defined terms involving products of singular distributions. The resulting junction conditions are collected in  \eqref{eq:JCbneq0} and are valid for the bi-scalar case $\alpha\neq0$ and for the pure-Ricci theory $\alpha=0$. In the first case, smooth matching at $\Sigma$ always entails the continuity of $R$ and $\mathcal{H}$, while a discontinuous $K_{\mu\nu}$ is only admissible in exceptional cases in which $\Mp^2+4\beta R^\sSigma=0$. In the second case, the conditions are the same except for the continuity of $\mathcal{H}$, which is no longer necessary.

    \item We have also provided details on the metric limit of the theory, namely, assuming that the torsion is vanishing from the beginning. Since the resulting theory is a particular case of metric $f(R)$ with $f'''(R)=0$, we have compared our results with those in the existing literature, \cite{Senovilla:2013vra}, finding they are in complete agreement with each other.
\end{itemize}

The obtained junction conditions could be used to explore physically relevant situations, such as spherical stellar collapse or brane-world scenarios, which can be used to understand better the dynamics of the bi-scalar theory. We will undertake these issues in a future work. 

\newpage

\acknowledgments{
The authors would like to thank J. M. M. Senovilla for useful discussions and feedback. The computations have been checked with xAct \cite{xAct}, a Mathematica package for tensorial symbolic calculus. This research was supported by the European Regional Development Fund through the Center of Excellence TK133 “The Dark Side of the Universe”. AJC was also supported by the Mobilitas Pluss post-doctoral grant MOBJD1035. FJMT is supported by the ``Fundaci\'on Ram\'on Areces''. ACT is supported by a Universidad Complutense de Madrid-Banco Santander predoctoral contract CT63/19-CT64/19, as well as a Univesidad Complutense short-term mobility grant EB25/22. ACT would like to express his most sincere gratitude to all the members of the Laboratory of Theoretical Physics of the University of Tartu, for their kind help and hospitality during his three-month research stay in Estonia, which facilitated the completion of this work. AdlCD acknowledges support from NRF grants no.120390, reference:BSFP190416431035; no.120396, reference:CSRP190405427545; projects PID2019- 108655GB-I00 and PID2021-122938NB-I00, MICINN Spain and BG20/00236 action, M. Universidades Spain.
}

\appendix

\section{Regular and singular parts. Useful formulae}
\label{app:regsing}

\subsection{First derivative of a distribution with singular part of the lowest order}

Let us focus on a tensor distribution whose singular part is purely monopolar, i.e., \begin{equation}
    \udis{F}_A  = \regpart\left[\udis{F}_A \right] + \singpart{0}\left[\udis{F}_A \right], \label{eq:decompS}
\end{equation}
where each of the pieces are, by definition, of the form
\begin{align}
    \regpart\left[\udis{F}_A \right] &= F^{+}_A \udis{\Theta}^{+} + F^{-}_A \udis{\Theta}^{-} \,,\\
   \singpart{0}\left[\udis{F}_A \right]&= F^\szero_A \del\,.
\end{align}

According to \eqref{eq:schemeD}, for the first derivative  we will have:
\begin{align}
    \disLCD_\mu \udis{F}_A  &= \regpart\left[\disLCD_\mu \udis{F}_A \right] + \singpart{0}\left[\disLCD_\mu \udis{F}_A \right]+ \singpart{1}\left[\disLCD_\mu \udis{F}_A \right]\,.
\end{align}
The decomposition is given by:
\begin{align}
    \regpart\left[\disLCD_\mu \udis{F}_A \right] &= (\LCD_\mu F_A)^{+} \, \udis{\Theta}^{+} + (\LCD_\mu F_A)^{-} \,\udis{\Theta}^{-} \,, \label{eq:decomDS1}\\
   \singpart{0}\left[\disLCD_\mu \udis{F}_A \right]&= \left( \epsilon n_\mu \big(\jump{F_A }- K^\sSigma F^\szero_A\big)+h^\sigma{}_\mu \LCD_\sigma F^\szero_A\right) \del \,,\label{eq:decomDS2}\\
   \singpart{1}\left[\disLCD_\mu \udis{F}_A \right]&= \disLCD_\sigma \left(\epsilon F^\szero_A n_\mu n^\sigma\del\right)\,,\label{eq:decomDS3}
\end{align}
where, we recall, $\epsilon = \pm 1$ (see eq. \eqref{eq:defepsilon}). Now we collect some special cases that are particularly useful for our purposes.

~

\noindent\textbf{Derivative of the Levi-Civita Ricci scalar}

For the Levi-Civita Ricci scalar ($\udis{F}_A \to \udis{\LCR}$), we have in our convention that $F^\szero_A  \to -2\epsilon \jump{K}$, so the corresponding first distributional derivative,
\begin{equation}
    \disLCD_\mu \LCR = \regpart\left[\disLCD_\mu \LCR\right] + \singpart{0}\left[\disLCD_\mu \LCR \right]+ \singpart{1}\left[\disLCD_\mu \LCR\right]\,, 
\end{equation}
has the following decomposition:
 \begin{align}
    \regpart\left[\disLCD_\mu \LCR\right] &= (\LCD_\mu\LCR)^{+} \, \udis{\Theta}^{+} + (\LCD_\mu \LCR)^{-} \,\udis{\Theta}^{-} \,,\\
   \singpart{0}\left[\disLCD_\mu \LCR\right]&= \epsilon\left( n_\mu \big(\jump{\LCR} + 2\epsilon K^\sSigma \jump{K}\big)- 2h^\sigma{}_\mu \LCD_\sigma \jump{K}\right) \del \,, \label{eq:Delta0DR}\\
   \singpart{1}\left[\disLCD_\mu\LCR \right]&=-2 \disLCD_\sigma \left(\jump{K} n_\mu n^\sigma\del\right)\,.
\end{align}

~

\noindent\textbf{Second derivative of a regular one-form}

Consider a regular one-form distribution $\udis{V}_\rho$, namely one which can be written as
\begin{equation}
    \udis{V}_\rho =  \regpart\left[\udis{V}_\rho\right] =  V^{+}_\rho \, \udis{\Theta}^{+} + V^{-}_\rho \,\udis{\Theta}^{-}.
\end{equation}
For instance, the torsion trace and axial vectors, $T_\mu$ and $S_\mu$, are tensors of this type, so the following equations are valid for them. The first distributional derivative,
\begin{equation}
    \disLCD_\mu\udis{V}_\rho  =  (\LCD_\mu V_\rho)^{+} \, \udis{\Theta}^{+} + (\LCD_\mu V_\rho)^{-} \,\udis{\Theta}^{-} + \epsilon n_\mu \jump{V_\rho}\del, \label{eq:decompDV}
\end{equation}
is a distribution of the form \eqref{eq:decompS}, i.e., its singular part only contains a contribution of the type $\singpart{0}$. Therefore, to compute the second distributional derivative, we can use the general formulas \eqref{eq:decomDS1}-\eqref{eq:decomDS3} with the substitutions $\udis{F}_A  \to \disLCD_\nu \udis{V}_\rho$, $F^{\pm}_A \to \LCD_\nu V^{\pm}_\rho$ and $F^\szero_A \to \epsilon n_\nu \jump{V_\rho}$. The result reads
 \begin{align}
    \regpart\left[\disLCD_\mu  \disLCD_\nu \udis{V}_\rho\right] &= (\LCD_\mu\LCD_\nu V_\rho)^{+} \, \udis{\Theta}^{+} + (\LCD_\mu\LCD_\nu V_\rho)^{-} \,\udis{\Theta}^{-} \,,\label{eq:regDDV}\\
   \singpart{0}\left[\disLCD_\mu  \disLCD_\nu \udis{V}_\rho\right]&= 
\epsilon\left[ n_\mu \left(\jump{\LCD_\nu V_\rho}- \epsilon K^\sSigma  n_\nu \jump{V_\rho}\right)+K^\sSigma_{\mu\nu} \jump{V_\rho}+n_\nu h^\sigma{}_\mu \LCD_\sigma   \jump{V_\rho}\right] \del   
   \,,\label{eq:Delta0DDV}\\
   \singpart{1}\left[\disLCD_\mu  \disLCD_\nu \udis{V}_\rho\right]&= \disLCD_\sigma \left(\jump{V_\rho} n_\mu n_\nu  n^\sigma\del\right)\,.\label{eq:Delta1DDV}
\end{align}
The second one can be further expanded by using \eqref{eq:jumpDV}.

\subsection{First and second derivative of a regular distribution}

In the equation of the metric \eqref{eq:EoM3}, terms of the form $\LCD_\mu \LCD_\nu \LCR$ and $\LCD_\mu \LCD_\nu \LCD_\rho T^\rho$ are present, and, according to the previous subsection, they would require one to study a structure that goes up to singular terms of the type $\singpart{2}$. However, thanks to the consistency conditions in Section \ref{cons:cond}, these two terms always appear in such a way that both $\LCR$ and $\LCD_\rho T^\rho$ (which, in principle, contain up to $\singpart{0}$) are purely regular. Here, we collect the results for the decomposition of the first and second derivatives of regular distributions, which are relevant for the analysis of those terms.

We start by considering a regular (but not necessarily continuous) tensor distribution:
\begin{equation}
    \udis{F}_A  = \regpart\left[\udis{F}_A \right] = F^{+}_A \udis{\Theta}^{+} + F^{-}_A \udis{\Theta}^{-} . \label{eq:decompR}
\end{equation}
This is a particular case of \eqref{eq:decompS} when $F^\szero_A=0$. In practice, we are interested in scalar distributions $\udis{F}_A \to \udis{F}$, since $\LCR$ and  $\LCD_\rho T^\rho$ are scalars, but we provide the expressions for objects with arbitrary tensor indices for the sake of completeness. 

The first and second derivatives of \eqref{eq:decompR} will have the following singular structure (see \eqref{eq:schemeD}):
\begin{align}
    \disLCD_\mu \udis{F}_A  &= \regpart\left[\disLCD_\mu \udis{F}_A \right] + \singpart{0}\left[\disLCD_\mu \udis{F}_A \right]\,, \\   
    \disLCD_\mu \disLCD_\nu \udis{F}_A  &= \regpart\left[\disLCD_\mu\disLCD_\nu \udis{F}_A \right] + \singpart{0}\left[\disLCD_\mu\disLCD_\nu \udis{F}_A \right]+ \singpart{1}\left[\disLCD_\mu \disLCD_\nu \udis{F}_A \right]\,.
\end{align}
For the first distributional derivative, the decomposition is simply \eqref{eq:decomDS1}-\eqref{eq:decomDS2} for $F^\szero_A=0$, whereas for the second derivative we get
\begin{align}
    \regpart\left[\disLCD_\mu\disLCD_\nu \udis{F}_A \right] &= (\LCD_\mu\LCD_\nu  F_A)^{+} \, \udis{\Theta}^{+} + (\LCD_\mu \LCD_\nu F_A)^{-} \,\udis{\Theta}^{-} \,, \label{eq:decomDDReg1}\\
   \singpart{0}\left[\disLCD_\mu\disLCD_\nu \udis{F}_A \right]&=  \epsilon \left[n_\mu \left(\jump{\LCD_\nu F_A}-\epsilon  K^\sSigma n_\nu\jump{F_A}\right) +K^\sSigma_{\mu\nu}\jump{F_A} + n_\nu h^\sigma{}_\mu \LCD_\sigma\jump{F_A}\right] \del \,,\label{eq:decomDDReg2}\\
   \singpart{1}\left[\disLCD_\mu\disLCD_\nu \udis{F}_A \right]&=   \disLCD_\sigma \left( \jump{F_A} n_\mu n_\nu n^\sigma \del\right) \,.\label{eq:decomDDReg3}
\end{align}
Observe that \eqref{eq:regDDV}-\eqref{eq:Delta1DDV} are particular cases of these expressions. Another interesting property that we can read from \eqref{eq:decomDDReg3} is that, for regular $\udis{F}_A$,
\begin{equation}
\singpart{1}\left[\disLCD_\mu\disLCD_\nu \udis{F}_A \right] =   \epsilon\disLCD_\sigma \left( \singpart{0}\left[\disLCD_{(\mu|} \udis{F}_{A} \right] n_{|\nu)} n^\sigma \right)\,.
\end{equation}

Below we provide the decomposition of the singular parts of the highest-order derivative terms in the equation of the metric \eqref{eq:EoM3}.

~

\noindent\textbf{Singular parts of the second derivative of the (regularized) Ricci scalar}

If we assume $\beta \jump{K}=0$, then the singular part of the combination $\beta\disLCD_\mu\disLCD_\nu \udis{\LCR}$ can be decomposed as follows into two parts:
\begin{align}
   \singpart{0}\left[\beta\disLCD_\mu\disLCD_\nu \udis{\LCR} \right]&=  \epsilon \beta \left[n_\mu \left(\partial_\nu\jump{\LCR}-\epsilon  K^\sSigma n_\nu\jump{\LCR}\right) +K^\sSigma_{\mu\nu}\jump{\LCR} + n_\nu h^\sigma{}_\mu \partial_\sigma\jump{\LCR}\right] \del \,,\label{eq:decomDDR1}\\
   \singpart{1}\left[\beta\disLCD_\mu\disLCD_\nu \udis{\LCR} \right]&=  \beta\disLCD_\sigma \left( \jump{\LCR} n_\mu n_\nu n^\sigma \del\right) \,,\label{eq:decomDDR2}
\end{align}
where we have used \eqref{eq:jumpDS}. In particular, if we contract with the metric, we get:
\begin{align}
   \singpart{0}\left[\beta\disLCD_\nu\disLCD^\nu \udis{\LCR} \right]&=  \epsilon \beta n^\nu \partial_\nu \jump{\LCR}  \del \,,\label{eq:decomtrDDR1}\\
   \singpart{1}\left[\beta\disLCD_\nu\disLCD^\nu \udis{\LCR} \right]&=  \epsilon\beta\disLCD_\sigma \left( \jump{\LCR} n^\sigma \del\right) \,.\label{eq:decomtrDDR2}
\end{align}

~

\noindent\textbf{Singular parts of the second derivative of the (regularized) divergence of the torsion trace}

For the singular part of $\beta\disLCD_\mu\disLCD_\nu \disLCD_\rho\udis{T}^\rho$ we get
\begin{align}
   \singpart{0}\left[\beta\disLCD_\mu\disLCD_\nu \disLCD_\rho\udis{T}^\rho \right]&=  \epsilon \beta\Big[n_\mu \left(\partial_\nu\jump{ \LCD_\rho T^\rho}-\epsilon  K^\sSigma n_\nu\jump{\LCD_\rho T^\rho}\right) \nonumber\\ 
   &\quad\qquad +K^\sSigma_{\mu\nu}\jump{\LCD_\rho T^\rho} + n_\nu h^\sigma{}_\mu \LCD_\sigma\jump{\LCD_\rho T^\rho}\Big] \del \,,\label{eq:decomDDDT1}\\
   \singpart{1}\left[\beta\disLCD_\mu\disLCD_\nu \disLCD_\rho\udis{T}^\rho \right]&=  \beta \disLCD_\sigma \left( \jump{\LCD_\rho T^\rho} n_\mu n_\nu n^\sigma \del\right) \,,\label{eq:decomDDDT2}
\end{align}
whose traces are:
\begin{align}
   \singpart{0}\left[\beta\disLCD_\nu\disLCD^\nu \disLCD_\rho\udis{T}^\rho \right]&=  \epsilon \beta n^\nu \partial_\nu\jump{ \LCD_\rho T^\rho}  \del \,,\label{eq:decomtrDDDT1}\\
   \singpart{1}\left[\beta\disLCD_\nu\disLCD^\nu \disLCD_\rho\udis{T}^\rho \right]&=  \epsilon\beta \disLCD_\sigma \left( \jump{\LCD_\rho T^\rho} n^\sigma \del\right) \,.\label{eq:decomtrDDDT2}
\end{align}
When using these expressions, it is useful to keep in mind that, from \eqref{eq:jumpDV}, and under $\beta \jump{K}=0$ and $\beta \jump{T_\mu}=0$, the following expression holds
\begin{equation}
    \beta\jump{\LCD_\rho T^\rho}= \epsilon \beta n^\lambda n^\rho \jump{\LCD_\lambda T_\rho} \,. \label{eq:jumpBetaDivT}
\end{equation}

\section{On the derivation of the equations of motion}
\label{app:EoM}

\subsection{Variations of the gravitational sector}

Let us first introduce the following notation for the variations of the gravitational part of the action:
\begin{align} 
    \mathcal{E}(T)^\mu &\equiv \frac{1}{\sqrt{|g|}}\frac{\delta\mathcal{S}_\text{g}}{\delta T_\mu}\,,
    &
    \mathcal{E}(S)^\mu &\equiv \frac{1}{\sqrt{|g|}}\frac{\delta\mathcal{S}_\text{g}}{\delta S_\mu}\,, \nonumber\\
    \mathcal{E}(t)_\sigma{}^{\mu\nu} &\equiv \frac{1}{\sqrt{|g|}} \frac{\delta\mathcal{S}_\text{g}}{\delta t^\sigma{}_{\mu\nu}} \,,
    &
    \mathcal{E}(g)_{\mu\nu} &\equiv \frac{1}{\sqrt{|g|}}\frac{\delta\mathcal{S}_\text{g}}{\delta g^{\mu\nu}}\,. \label{eq:var0}
\end{align}
Such variations are given by:
\begin{align} 
    \mathcal{E}(T)^\mu&=\left(M_T^2-\frac{8\beta}{3} R\right)T^\mu +\frac{4\alpha}{3}\mathcal{H}S^\mu -4\beta \LCD^\mu R \,, \label{eq:varT0}\\
    \mathcal{E}(S)^\mu&= \left(M_S^2+\frac{\beta}{6} R\right)S^\mu +\frac{4\alpha}{3}\mathcal{H}T^\mu +2\alpha \LCD^\mu \mathcal{H}   \,, \label{eq:varS0}\\
    \mathcal{E}(t)_\sigma{}^{\mu\nu}&= \left(M_t^2+2\beta R\right)t_\sigma{}^{\mu\nu} + 2\alpha\mathcal{H} \LCten^{\mu\nu\rho\lambda}t_{\sigma \rho\lambda} \,, \label{eq:vart0}\\
    \mathcal{E}(g)_{\mu\nu}&=
    \frac{\Mp^2}{2} \left(E_{\mu\nu}-\frac{1}{2}g_{\mu\nu}E_\rho{}^\rho\right)-2\beta (\LCD_{\mu}\LCD_{\nu}- g_{\mu\nu}\LCbox) R  + P_{\mu\nu}\,, \label{eq:varg0}
\end{align}
where $M_T$ and $M_S$ are defined in \eqref{eq:defMSMT}, 
\begin{equation}
    M_t^2 \equiv m_t^2 + \frac{\Mp^2}{2}\,,
\end{equation}
and
\begin{align}
   \Mp^2 E_{\mu\nu}&\equiv \Mp^2\mathring{R}_{\mu\nu} + 4\beta \LCR_{\mu\nu}R + 4 \alpha S_{(\mu}\LCD_{\nu)}\mathcal{H}- 8\beta T_{(\mu}\LCD_{\nu)}R \nonumber \\
     &
    \quad + \left(M_S^2+ \frac{\beta}{6} R\right)S_\mu S_\nu + \frac{8}{3}\alpha \mathcal{H} S_{(\mu} T_{\nu)} + \left(M_T^2- \frac{8\beta}{3} R\right)T_\mu T_\nu\nonumber \\
     &
    \quad -g_{\mu\nu}(\alpha \mathcal{H}^2 + \beta R^2)\,,\label{eq:deftensorE}   \\ 
   P_{\mu\nu}&\equiv \left(\frac{M_t^2}{2} + \beta R\right)\left(2t^{\rho\sigma}{}_\mu t_{\rho\sigma\nu}-t_\mu{}^{\rho\sigma}t_{\nu\rho\sigma} -\frac{1}{2}g_{\mu\nu} t^{\rho\lambda\sigma}t_{\rho\lambda\sigma}   
   \right)\nonumber\\
   &\quad - \alpha\mathcal{H}  \left[
   \LCten_{\rho\lambda\sigma\tau}(t_{\mu}{}^{\rho\lambda}t_{\nu}{}^{\sigma\tau}+g_{\mu\nu}t_{\alpha}{}^{\rho\lambda}t^{\alpha\sigma\tau})+ 4 t_{\alpha}{}^{\rho\lambda}t^{\alpha\sigma}{}_{(\mu} \LCten_{\nu)\sigma\rho\lambda}
   \right]\,.\label{eq:deftensorP}
\end{align}
Note that $P_{\mu\nu}$ vanishes when $t^\sigma{}_{\mu\nu}=0$.\footnote{To find the expression \eqref{eq:deftensorP} with xAct, we had to perform some simplifications based on the following fact. For any tensor $Z_{\mu\nu\rho\lambda}$, with the same symmetries as the Riemann tensor (antisymmetry in both pairs and symmetry under pair exchange), it can be shown that, in four dimensions, $Z_{\mu[\nu\rho\lambda]}= Z_{[\mu\nu\rho\lambda]}$, which leads to the useful identity:
\[
Z^{\mu[\sigma\rho\lambda]}Z_{\nu[\sigma\rho\lambda]}=\frac{1}{4}\delta^\mu_{\nu}Z^{\tau\sigma\rho\lambda}Z_{\tau[\sigma\rho\lambda]}.
\]
We rearranged it as 
\[
Z_\mu{}^{\lambda\sigma\rho}Z_{\nu\sigma\rho\lambda} = -\frac{1}{2}Z_\mu{}^{\sigma\rho\lambda}Z_{\nu\sigma\rho\lambda}+\frac{3}{8}g_{\mu\nu}Z^{\tau\sigma\rho\lambda}Z_{\tau[\sigma\rho\lambda]}
\]
and use it for $Z_{\mu\nu\rho\lambda} = t_{\alpha\mu\nu}t^{\alpha}{}_{\rho\lambda}$.}

\subsection{On the equation of motion of the tensor part of the torsion}
\label{app:EoMt}

In this section, we are going to analyze the equation of $t^\sigma{}_{\mu\nu}$. Ignoring the matter contribution to it, we simply get from \eqref{eq:vart0}:
\begin{equation}
    0=\mathcal{E}(t)_{\sigma\mu\nu}= C_1 t_{\sigma\mu\nu} + C_2 \LCten_{\mu\nu}{}^{\rho\lambda}t_{\sigma \rho\lambda}\,,
\end{equation}
with $C_1 \equiv M_t^2+2\beta R$ and $C_2\equiv2\alpha\mathcal{H}$. First, we notice that, if $C_1=0$ and $C_2\neq 0$, or if $C_1\neq0$ and $C_2= 0$, we automatically have $t_{\sigma \mu\nu}=0$. If both are non-vanishing, then one can construct the combination:
\begin{equation}
    0=C_1 \mathcal{E}(t)_{\sigma\mu\nu}- C_2 \LCten_{\mu\nu}{}^{\alpha\beta}\mathcal{E}(t)_{\sigma\alpha\beta} = (C_1^2+ 4 C_2^2)t_{\sigma \mu\nu},
\end{equation}
which leads to $t_{\sigma \mu\nu}=0$. Finally, in the case in which both $C_1$ and $C_2$ are vanishing,
\begin{equation}
    \beta R= - \frac{M_t^2}{2}, \mybigskip \alpha\mathcal{H}=0\,. \label{eq:condC1C2}
\end{equation}
Under conditions \eqref{eq:condC1C2}, the tensor $t^\sigma{}_{\mu\nu}$ remains (dynamically) undetermined and only subjected to the previous restrictions. Indeed, it disappears from the rest of the equations (see e.g. that $P_{\mu\nu}$ in \eqref{eq:deftensorP} vanishes identically under \eqref{eq:condC1C2}). The remaining equations of motion, after introducing the matter currents \eqref{eq:defLJT} and making use of \eqref{eq:condC1C2}, become
\begin{align} 
    L_\mu&= \left(M_T^2+\frac{4}{3}M_t^2\right) T_\mu \quad\  = \left(m_T^2+\frac{4}{3}m_t^2\right) T_\mu \,,\\
   J_\mu&= \left(M_S^2-\frac{1}{12}M_t^2 \right)S_\mu \quad = \left(m_S^2-\frac{1}{12}m_t^2 \right)S_\mu    \,,\\
    T_{\mu\nu}&=
    \left(\Mp^2-2M_t^2\right)\mathring{G}_{\mu\nu}  -\frac{M_t^2}{2}g_{\mu\nu}R  \nonumber\\
    &\quad +  \frac{1}{2}\left[ 2J_{(\mu} S_{\nu)}+ 2L_{(\mu} T_{\nu)} -g_{\mu\nu}(J_\lambda S^\lambda+L_\lambda T^\lambda)\right]\,.
\end{align}
We see that the dynamics of the two vector variables is also absent and that, in vacuum, they are vanishing for generic values of the parameters (except those in the discrete set making the combinations in brackets vanish). For $\beta=0$, we get $M_t=0$, corresponding to GR sourced by an effective energy-momentum tensor, i.e., to $\mathring{G}_{\mu\nu} = \Mp^{-2} T_{\mu\nu}^\text{eff}$, with
\begin{equation}
    T_{\mu\nu}^\text{eff}\equiv T_{\mu\nu} -  \frac{1}{2}\left[ 2J_{(\mu} S_{\nu)}+ 2L_{(\mu} T_{\nu)} -g_{\mu\nu}(J_\lambda S^\lambda+L_\lambda T^\lambda)\right]\,.
\end{equation}
If $\beta\neq 0$, we also get an effective cosmological constant and a correction to Newton's constant:
\begin{align}
    \left(\Mp^2-2M_t^2\right)\mathring{G}_{\mu\nu}  +\frac{M_t^4}{4\beta}g_{\mu\nu}  = T_{\mu\nu}^\text{eff}\,.
\end{align}
In both cases, the tensor part of the torsion drops from the whole set of equations of motion.

From the previous calculation,s we can state that the only way for the torsion tensor to contribute dynamically to the field equations is by having a non-minimal derivative coupling with matter. Otherwise, it can only contribute as a term that can be absorbed in the energy-momentum tensor. Therefore, we could assume that all the dynamics would be encoded in the axial and trace vectors and omit the tensorial part from the beginning. This agrees with the considerations in \cite{BeltranJimenez:2019hrm}, where the authors justify this choice by performing a Legendre transformation on the action of the theory.

\section{Subtle aspects regarding  the prescription for distributional promotion}
\label{app:subtleties}



For equations involving products of two fields, one corresponding to a regular distribution $\udis{X}$, and another one, $\udis{Y}$, having a non-trivial monopolar part, our prescription \eqref{eq:prescTheDel} would be:
\begin{equation}
    X Y \quad \overset{\text{def}}{\longrightarrow}  \quad \udis{X}\,\udis{Y} \equiv X^+ Y^+ \udis{\Theta}^+ \ +\ X^-Y^-\udis{\Theta}^- \ +\  X^\sSigma \singpart{0}[Y]\,. 
\end{equation}
This is the same result we would have obtained through the naive promotion
\begin{equation}
     XY \quad \overset{\text{naive}}{\longrightarrow}  \quad  \regpart [X] (\regpart[Y] + \singpart{0}[Y])\,,
\end{equation}
as well as the identifications    
\begin{equation}           
    \udis{\Theta}^\pm\udis{\Theta}^\pm =\udis{\Theta}^\pm\,,\myskip \udis{\Theta}^\pm\udis{\Theta}^\mp=0\,,\myskip \udis{\Theta}^\pm \del= \frac{1}{2}\del\,.\label{eq:identif}
\end{equation}
However, the equations of bi-scalar Poincaré gravity involve products of three objects, two of which are regular and the remaining one monopolar (see, e.g. the term $    S^\sigma T^{\lambda} \LCD_\sigma S_{\lambda}$ in \eqref{eq:Tmnsing}). In such a case, one might be tempted to think that the identifications in \eqref{eq:identif} would still work. However, there is a problem: such a product is not associative. Compare, for example,
\begin{align}
    (\udis{\Theta}^+\udis{\Theta}^+)\del &= \udis{\Theta}^+\del = \frac{1}{2}\del \,,\nonumber\\
     \udis{\Theta}^+(\udis{\Theta}^+\del) &= \frac{1}{2}\udis{\Theta}^+\del = \frac{1}{4}\del \,.        
\end{align}
Thus, one is forced to make a choice. Our prescription \eqref{eq:prescTheDel} corresponds to ``multiplying the $\udis{\Theta}$'s by the $\del$ one by one and not among themselves''. The other choice would have led to    
\begin{equation}
        X_1 X_2 Y \quad \overset{\text{other def.}}{\longrightarrow}  \quad  X^+_1 X^+_2 Y^+ \udis{\Theta}^+\ +\ X^-_1 X^-_2 Y^-\udis{\Theta}^-\ +\ (X_1 X_2)^\sSigma \singpart{0}[Y]\,. 
\end{equation}
Notice that the last term is different from the one obtained from \eqref{eq:prescTheDel}, which is $X^\sSigma_1 X^\sSigma_2\singpart{0}[Y]$.\footnote{
    Observe that the ambiguity is even worse when there are more than two $X_i$. For example, for three of them, there are five possibilities: $(X_1 X_2)^\sSigma X^\sSigma_3$,  $(X_3 X_2)^\sSigma X^\sSigma_1$, $(X_1 X_3)^\sSigma X^\sSigma_2$, $(X_1 X_2 X_3)^\sSigma$ and $X^\sSigma_1X^\sSigma_2X^\sSigma_3$. However, to reproduce the results in this paper, only the case of two $X_i$ will be relevant.
    }
    
Finally, let us justify why our choice is \emph{sensible}, by defining it in a different way. The underlying idea is that we can treat the $\udis{\Theta}^\pm$ inside the $\udis{X}_i$ simply as the functions $\Theta^\pm$. In other words, we are identifying the distributions $\udis{X}_i$ with the associated tensors $X_i$. Intuitively, this makes sense, since regular tensor distributions do not present singular contributions anywhere (of course, this is not mathematically rigorous, as $\udis{X}_i$ and  $X_i$ belong to different spaces). This leads us to define the product $\udis{X}_1 \udis{X}_2 \udis{Y}$ as the distribution acting on test functions $\varphi$ as follows:
 \begin{align}
    \langle \udis{X}_1 \udis{X}_2 \udis{Y},\, \varphi \rangle &\equiv \langle X_1 X_2 \udis{Y},\, \varphi \rangle\nonumber\\
        &= \langle  \udis{Y},\, X_1 X_2 \varphi \rangle\nonumber\\
        &= \langle  \regpart[Y],\, X_1 X_2 \varphi \rangle + \langle  \singpart{0}[Y],\, X_1 X_2 \varphi \rangle\nonumber\\
        &= \langle  \udis{\Theta}^+,\, X_1 X_2 Y^+ \varphi \rangle + \langle  \udis{\Theta}^-,\, X_1 X_2 Y^- \varphi \rangle+ \langle  \singpart{0}[Y],\, (X_1 X_2 \varphi)|_\Sigma \rangle\nonumber\\
        &= \Big\langle  (X^+_1 X^+_2 Y^+)\udis{\Theta}^+,\,  \varphi \Big\rangle + \Big\langle  (X^-_1 X^-_2 Y^-)\udis{\Theta}^-,\, \varphi \Big\rangle + \Big\langle  \singpart{0}[Y],\, X_1|_\Sigma X_2|_\Sigma \varphi \Big\rangle\nonumber\\
        &= \Big\langle  \Big[(X^+_1 X^+_2 Y^+)\udis{\Theta}^+ + (X^-_1 X^-_2 Y^-)\udis{\Theta}^- + X^\sSigma_1 X^\sSigma_2\singpart{0}[Y]\Big],  \,\varphi \Big\rangle \,.\label{eq:XXYint}
\end{align}
This is compatible with our prescription \eqref{eq:prescTheDel}.     In the last step, we have used $X_i|_\Sigma = X^\sSigma_i$, which holds for cases relevant to the computations in this work: $X_i = T_\mu, S_\mu$. Note that (except for the definition in the first line) all the steps in \eqref{eq:XXYint} are mathematically rigorous.

\section{Singular contributions to the equations of motion}
\label{app:tablesterms}
In the first column of the following tables, we show the different types of terms as they appear in the equations of motion (with $t^\rho{}_{\mu\nu}=0$) \eqref{eq:EoM1}-\eqref{eq:EoM3}. In the second column, we present the terms arising from the aforementioned terms in the first column when expressed in terms of basic variables and their derivatives (including Levi-Civita curvatures). Then, in the third column, we show both the regular and singular contributions obtained from the previous terms when promoted to distributions while subject to the preliminary conditions \eqref{eq:condNoSingTSg}-\eqref{eq:condnojumpg}. Finally, in the last column we display the  terms that survive the enforcement of all the consistency conditions \eqref{eq:condNoSingTSg}-\eqref{eq:condJumpnS}. Indices of the vectors have been omitted for clarity almost everywhere in the following tables, except for the combination $\alpha\LCD_\mu S^\mu$, in order to clearly signal where condition \eqref{eq:condJumpnS} has been used. Finally, we clarify that by ``undefined as distribution'' we mean that those terms would contain powers of singular parts. Recall that, although the products of $\udis{\Theta}\,\udis{\Theta}$ and $\udis{\Theta}\del$ would also be mathematically undefined, our prescription \eqref{eq:prescTheDel} allows to deal with them. 

\begin{table}[ht]
    \centering
    \renewcommand\arraystretch{1.2}
    \begin{tabular}{ | m{2.5cm}| m{3.5cm} | m{3.5cm}| m{3.5cm} | } 
    \hline
    {\footnotesize Terms in \eqref{eq:EoM1}} & {\footnotesize In terms of basic fields} & {\footnotesize Pieces after \eqref{eq:condNoSingTSg}-\eqref{eq:condnojumpg}} & {\footnotesize Pieces after \eqref{eq:condNoSingTSg}-\eqref{eq:condJumpnS}}\\
    \hline
    \hline
    $M_T^2 T$&  $M_T^2 T$     & $\regpart$  & $\regpart$  \\ 
    \hline
    \multirow{3}{*}{$\beta T R $} 
    &$\beta T \LCR$           & $\regpart + \singpart{0}$  & $\regpart$ \\ 
    &$\beta T \LCD T$         & $\regpart + \singpart{0}$  & $\regpart$ \\ 
    &$\beta SST, \beta TTT$   & $\regpart$                 & $\regpart$ {\footnotesize (continuous)} \\
    \hline
    \multirow{2}{*}{$\alpha S\mathcal{H} $}
    &$\alpha S\LCD_\mu S^\mu$ & $\regpart + \singpart{0}$  & $\regpart$ \\ 
    &$\alpha TSS$             & $\regpart $                & $\regpart$ \\ 
    \hline
    \multirow{3}{*}{$\beta \LCD R$} 
    &$\beta \LCD \LCR$        & $\regpart + \singpart{0}+ \singpart{1}$  & $\regpart+ \singpart{0}$ \\ 
    &$\beta \LCD \LCD T$      & $\regpart + \singpart{0}+ \singpart{1}$  & $\regpart+ \singpart{0}$ \\ 
    &$\beta S\LCD S$, $\beta T\LCD T$          & $\regpart + \singpart{0}$  & $\regpart$ \\
    \hline
\end{tabular}
    \renewcommand\arraystretch{1}
    \caption{Terms in the equation of $T_\mu$, \eqref{eq:EoM1}, where we have indicated the regular and singular contributions to the different terms involved in the equation.}
    \label{tab:termsEqT}
\end{table}

\begin{table}[ht]
    \centering
    \renewcommand\arraystretch{1.2}
    \begin{tabular}{ | m{2.5cm}| m{3.5cm} | m{3.5cm}| m{3.5cm} | } 
    \hline
    {\footnotesize Terms in \eqref{eq:EoM2}} & {\footnotesize In terms of basic fields} & {\footnotesize Pieces after \eqref{eq:condNoSingTSg}-\eqref{eq:condnojumpg}} & {\footnotesize Pieces after \eqref{eq:condNoSingTSg}-\eqref{eq:condJumpnS}}\\
    \hline
    \hline
    $M_S^2 S$&  $M_S^2 S$       & $\regpart$  & $\regpart$  \\ 
    \hline
    \multirow{3}{*}{$\beta SR $} 
    &$\beta S \LCR$             & $\regpart + \singpart{0}$  & $\regpart$ \\ 
    &$\beta S \LCD T$           & $\regpart + \singpart{0}$  & $\regpart$ \\ 
    &$\beta SSS, \beta STT$     & $\regpart$                 & $\regpart$ {\footnotesize (continuous)} \\
    \hline
    \multirow{2}{*}{$\alpha T\mathcal{H}$}
    &$\alpha T\LCD_\mu S^\mu$   & $\regpart + \singpart{0}$  & $\regpart$ \\ 
    &$\alpha STT$               & $\regpart $                & $\regpart$ \\ 
    \hline
    \multirow{2}{*}{$\alpha \LCD \mathcal{H}$} 
    &$\alpha\LCD(\LCD_\mu S^\mu)$ & $\regpart + \singpart{0}+ \singpart{1}$  & $\regpart+ \singpart{0}$ \\ 
    &$\alpha T\LCD S$, $\alpha S\LCD T$           & $\regpart + \singpart{0}$  & $\regpart + \singpart{0}$ \\
    \hline
\end{tabular}
    \renewcommand\arraystretch{1}
    \caption{Terms in the equation of $S_\mu$, \eqref{eq:EoM2}, indicating the regular and singular parts of the different terms.}
    \label{tab:termsEqS}
\end{table}

\begin{table}[ht]
    \centering
    \renewcommand\arraystretch{1.2}
    \begin{tabular}{ | m{2.5cm}| m{3.5cm} | m{3.5cm}| m{3.5cm} | } 
    \hline
    {\footnotesize Terms in \eqref{eq:EoM3}} & {\footnotesize In terms of basic fields} & {\footnotesize Pieces after \eqref{eq:condNoSingTSg}-\eqref{eq:condnojumpg}} & {\footnotesize Pieces after \eqref{eq:condNoSingTSg}-\eqref{eq:condJumpnS}}\\
    \hline
    \hline
    $M_S^2 SS$, $M_T^2 TT$&  $M_S^2 SS$, $M_T^2 TT$& $\regpart$  & $\regpart$  \\ 
    \hline
    $\Mp^2\LCR_{\mu\nu}$&  $\Mp^2\LCR_{\mu\nu}$& $\regpart+ \singpart{0}$  & $\regpart+ \singpart{0}$  \\ 
    \hline
    \multirow{3}{*}{$\beta\LCR_{\mu\nu} R$} 
    &$\beta\LCR_{\mu\nu}\LCR$     & {\footnotesize undefined as distribution}  & $\regpart+ \singpart{0}$ \\ 
    &$\beta\LCR_{\mu\nu} \LCD T$  & {\footnotesize undefined as distribution}  & $\regpart$ \\ 
    &$\beta\LCR_{\mu\nu} TT$, $\beta\LCR_{\mu\nu} SS$   & $\regpart+ \singpart{0}$                 & $\regpart+ \singpart{0}$ \\
    \hline
    \multirow{3}{*}{$\beta TT R $} 
    &$\beta TT \LCR$           & $\regpart + \singpart{0}$  & $\regpart$ \\ 
    &$\beta TT \LCD T$         & $\regpart + \singpart{0}$  & $\regpart$ \\ 
    &$\beta SSTT, \beta TTTT$   & $\regpart$                 & $\regpart$ {\footnotesize (continuous)} \\
    \hline
    \multirow{3}{*}{$\beta SS R $} 
    &$\beta SS \LCR$           & $\regpart + \singpart{0}$  & $\regpart$ \\ 
    &$\beta SS \LCD T$         & $\regpart + \singpart{0}$  & $\regpart$ \\ 
    &$\beta SSSS, \beta SSTT$   & $\regpart$                 & $\regpart$ {\footnotesize (continuous)} \\ 
    \hline
    \multirow{2}{*}{$\alpha TS\mathcal{H} $}
    &$\alpha TS\LCD_\mu S^\mu$ & $\regpart + \singpart{0}$  & $\regpart$ \\ 
    &$\alpha TTSS$             & $\regpart $                & $\regpart$ \\  
    \hline
    \multirow{3}{*}{$\beta T\LCD R$} 
    &$\beta T \LCD \LCR$        & $\regpart + \singpart{0}+ \singpart{1}$  & $\regpart+ \singpart{0}$ \\ 
    &$\beta T \LCD \LCD T$      & $\regpart + \singpart{0}+ \singpart{1}$  & $\regpart+ \singpart{0}$ \\ 
    &$\beta TS\LCD S$, $\beta TT\LCD T$          & $\regpart + \singpart{0}$  & $\regpart$ \\
    \hline
    \multirow{2}{*}{$\alpha S \LCD \mathcal{H}$} 
    &$\alpha S\LCD(\LCD_\mu S^\mu)$ & $\regpart + \singpart{0}+ \singpart{1}$  & $\regpart+ \singpart{0}$ \\ 
    &$\alpha TS\LCD S$, $\alpha SS\LCD T$           & $\regpart + \singpart{0}$  & $\regpart + \singpart{0}$ \\ 
    \hline
    \multirow{3}{*}{$\alpha \mathcal{H}^2$} 
    &$\alpha(\LCD_\mu S^\mu)^2$ & {\footnotesize undefined as distribution}  & $\regpart$ \\ 
    &$\alpha ST\LCD_\mu S^\mu$  & $\regpart + \singpart{0}$  & $\regpart$ \\
    &$\alpha SSTT$              & $\regpart$  & $\regpart$ \\
    \hline
    \multirow{6}{*}{$\beta R^2$} 
    &$\beta \LCR^2$                   & {\footnotesize undefined as distribution}  & $\regpart$ \\ 
    &$\beta \LCR\LCD T$               & {\footnotesize undefined as distribution}  & $\regpart$ \\ 
    &$\beta SS\LCR$, $\beta TT\LCR$    & $\regpart + \singpart{0}$  & $\regpart$ \\ 
    &$\beta \LCD T \LCD T$            & {\footnotesize undefined as distribution}  & $\regpart$ \\ 
    &$\beta SS\LCD T$, $\beta TT\LCD T$& $\regpart + \singpart{0}$  & $\regpart$ \\ 
    &$\beta \times \text{(powers of}\,T,S\text{)}$ &$\regpart$  & $\regpart$ {\footnotesize (continuous)} \\
    \hline
    \multirow{4}{*}{$\beta \LCD\LCD R$} 
    &$\beta \LCD\LCD R$                   & $\regpart + \singpart{0}+ \singpart{1}+ \singpart{2}$ & $\regpart + \singpart{0}+ \singpart{1}$ \\ 
    &$\beta \LCD\LCD\LCD T$               & $\regpart + \singpart{0}+ \singpart{1}+ \singpart{2}$ & $\regpart + \singpart{0}+ \singpart{1}$ \\ 
    &$\beta S\LCD\LCD S$, $\beta T \LCD\LCD T$    & $\regpart + \singpart{0}+ \singpart{1}$  & $\regpart+ \singpart{0}$ \\ 
    &$\beta \LCD T \LCD T$, $\beta \LCD S \LCD S$            & {\footnotesize undefined as distribution}  & $\regpart$ \\ 
    \hline
\end{tabular}
    \renewcommand\arraystretch{1}
    \caption{Terms in the equation of $g_{\mu\nu}$, \eqref{eq:EoM3}, including the regular and singular parts of the different terms.}
    \label{tab:termsEqMetric}
\end{table}

\newpage

~

\newpage

\bibliographystyle{JHEP}
\bibliography{references.bib}

\providecommand{\href}[2]{#2}\begingroup\raggedright\begin{thebibliography}{10}

\bibitem{Misner:1974qy}
C.~W. Misner, K.~Thorne and J.~Wheeler, \emph{{Gravitation}}.
\newblock W. H. Freeman, San Francisco, 1973.

\bibitem{https://doi.org/10.1002/andp.19243791403}
K.~Lanczos, \emph{Flächenhafte verteilung der materie in der einsteinschen
  gravitationstheorie},
  \href{http://dx.doi.org/https://doi.org/10.1002/andp.19243791403}{\emph{Annalen
  der Physik} {\bfseries 379} (1924) 518--540},
  [\href{https://arxiv.org/abs/https://onlinelibrary.wiley.com/doi/pdf/10.1002/andp.19243791403}{{\ttfamily
  https://onlinelibrary.wiley.com/doi/pdf/10.1002/andp.19243791403}}].

\bibitem{darmois1927equations}
G.~Darmois, \emph{Les {\'e}quations de la gravitation einsteinienne}.
\newblock Gauthier-Villars Paris, 1927.

\bibitem{Lichnerowicz:107002}
A.~Lichnerowicz, \emph{{Théories relativistes de la gravitation et de
  l'électromagnétisme: relativité générale et théories unitaires}}.
\newblock Masson, Paris, 1955.

\bibitem{o1952jump}
S.~O'Brien and J.~L. Synge, \emph{Jump conditions at discontinuities in general
  relativity}, {\emph{Commun. Dublin Inst. Adv. Stud., A} (1952) }.

\bibitem{bel1967conditions}
L.~Bel and A.~Hamoui, \emph{Les conditions de raccordement en relativit{\'e}
  g{\'e}n{\'e}rale},  in \emph{Annales de l'IHP Physique th{\'e}orique},
  pp.~229--244, 1967.

\bibitem{taub1980space}
A.~Taub, \emph{Spacetimes with distribution valued curvature tensors},
  \href{http://dx.doi.org/10.1063/1.524568}{\emph{Journal of Mathematical
  Physics} {\bfseries 21} (1980) 1423--1431}.

\bibitem{bonnor1981junction}
W.~Bonnor and P.~Vickers, \emph{Junction conditions in general relativity},
  \href{http://dx.doi.org/10.1007/BF00766295}{\emph{General Relativity and
  Gravitation} {\bfseries 13} (1981) 29--36}.

\bibitem{clarke1987junction}
C.~Clarke and T.~Dray, \emph{Junction conditions for null hypersurfaces},
  \href{http://dx.doi.org/10.1088/0264-9381/4/2/010}{\emph{Classical and
  Quantum Gravity} {\bfseries 4} (1987) 265}.

\bibitem{barrabes1989singular}
C.~Barrabes, \emph{Singular hypersurfaces in general relativity: a unified
  description},
  \href{http://dx.doi.org/10.1088/0264-9381/6/5/003}{\emph{Classical and
  Quantum Gravity} {\bfseries 6} (1989) 581}.

\bibitem{Mars:1993mj}
M.~Mars and J.~M. Senovilla, \emph{{Geometry of general hypersurfaces in
  space-time: Junction conditions}},
  \href{http://dx.doi.org/10.1088/0264-9381/10/9/026}{\emph{Class. Quant.
  Grav.} {\bfseries 10} (1993) 1865--1897},
  [\href{https://arxiv.org/abs/gr-qc/0201054}{{\ttfamily gr-qc/0201054}}].

\bibitem{Israel:1966rt}
W.~Israel, \emph{{Singular hypersurfaces and thin shells in general
  relativity}}, \href{http://dx.doi.org/10.1007/BF02710419}{\emph{Nuovo Cim. B}
  {\bfseries 44S10} (1966) 1}.

\bibitem{barrabes1991thin}
C.~Barrabes and W.~Israel, \emph{Thin shells in general relativity and
  cosmology: The lightlike limit},
  \href{http://dx.doi.org/10.1103/PhysRevD.43.1129}{\emph{Physical Review D}
  {\bfseries 43} (1991) 1129}.

\bibitem{Deruelle:2007pt}
N.~Deruelle, M.~Sasaki and Y.~Sendouda, \emph{{Junction conditions in f(R)
  theories of gravity}},
  \href{http://dx.doi.org/10.1143/PTP.119.237}{\emph{Prog. Theor. Phys.}
  {\bfseries 119} (2008) 237--251},
  [\href{https://arxiv.org/abs/0711.1150}{{\ttfamily 0711.1150}}].

\bibitem{Clifton:2012ry}
T.~Clifton, P.~Dunsby, R.~Goswami and A.~M. Nzioki, \emph{{On the absence of
  the usual weak-field limit, and the impossibility of embedding some known
  solutions for isolated masses in cosmologies with f(R) dark energy}},
  \href{http://dx.doi.org/10.1103/PhysRevD.87.063517}{\emph{Phys. Rev. D}
  {\bfseries 87} (2013) 063517},
  [\href{https://arxiv.org/abs/1210.0730}{{\ttfamily 1210.0730}}].

\bibitem{Senovilla:2013vra}
J.~M. Senovilla, \emph{{Junction conditions for F(R)-gravity and their
  consequences}},
  \href{http://dx.doi.org/10.1103/PhysRevD.88.064015}{\emph{Phys. Rev. D}
  {\bfseries 88} (2013) 064015},
  [\href{https://arxiv.org/abs/1303.1408}{{\ttfamily 1303.1408}}].

\bibitem{Casado-Turrion:2022xkl}
A.~Casado-Turri\'on, A.~de~la Cruz-Dombriz and A.~Dobado, \emph{{Is
  gravitational collapse possible in f(R) gravity?}},
  \href{http://dx.doi.org/10.1103/PhysRevD.105.084060}{\emph{Phys. Rev. D}
  {\bfseries 105} (2022) 084060},
  [\href{https://arxiv.org/abs/2202.04439}{{\ttfamily 2202.04439}}].

\bibitem{Vignolo:2018eco}
S.~Vignolo, R.~Cianci and S.~Carloni, \emph{{On the junction conditions in
  $f(R)$-gravity with torsion}},
  \href{http://dx.doi.org/10.1088/1361-6382/aab6fe}{\emph{Class. Quant. Grav.}
  {\bfseries 35} (2018) 095014},
  [\href{https://arxiv.org/abs/1801.08344}{{\ttfamily 1801.08344}}].

\bibitem{Olmo:2020fri}
G.~J. Olmo and D.~Rubiera-Garcia, \emph{{Junction conditions in Palatini $f(R)$
  gravity}}, \href{http://dx.doi.org/10.1088/1361-6382/abb924}{\emph{Class.
  Quant. Grav.} {\bfseries 37} (2020) 215002},
  [\href{https://arxiv.org/abs/2007.04065}{{\ttfamily 2007.04065}}].

\bibitem{Rosa:2021mln}
J.~L. Rosa and J.~P.~S. Lemos, \emph{{Junction conditions for generalized
  hybrid metric-Palatini gravity with applications}},
  \href{http://dx.doi.org/10.1103/PhysRevD.104.124076}{\emph{Phys. Rev. D}
  {\bfseries 104} (2021) 124076},
  [\href{https://arxiv.org/abs/2111.12109}{{\ttfamily 2111.12109}}].

\bibitem{Reina:2015gxa}
B.~Reina, J.~M.~M. Senovilla and R.~Vera, \emph{{Junction conditions in
  quadratic gravity: thin shells and double layers}},
  \href{http://dx.doi.org/10.1088/0264-9381/33/10/105008}{\emph{Class. Quant.
  Grav.} {\bfseries 33} (2016) 105008},
  [\href{https://arxiv.org/abs/1510.05515}{{\ttfamily 1510.05515}}].

\bibitem{Padilla:2012ze}
A.~Padilla and V.~Sivanesan, \emph{{Boundary Terms and Junction Conditions for
  Generalized Scalar-Tensor Theories}},
  \href{http://dx.doi.org/10.1007/JHEP08(2012)122}{\emph{JHEP} {\bfseries 08}
  (2012) 122}, [\href{https://arxiv.org/abs/1206.1258}{{\ttfamily 1206.1258}}].

\bibitem{delaCruz-Dombriz:2014zaa}
A.~de~la Cruz-Dombriz, P.~K.~S. Dunsby and D.~Saez-Gomez, \emph{{Junction
  conditions in extended Teleparallel gravities}},
  \href{http://dx.doi.org/10.1088/1475-7516/2014/12/048}{\emph{JCAP} {\bfseries
  12} (2014) 048}, [\href{https://arxiv.org/abs/1406.2334}{{\ttfamily
  1406.2334}}].

\bibitem{Arkuszewski:1975fz}
W.~Arkuszewski, W.~Kopczynski and V.~Ponomarev, \emph{{Matching Conditions in
  the Einstein-Cartan Theory of Gravitation}},
  \href{http://dx.doi.org/10.1007/BF01629248}{\emph{Commun. Math. Phys.}
  {\bfseries 45} (1975) 183--190}.

\bibitem{Macias:2002sr}
A.~Macias, C.~Lammerzahl and L.~O. Pimentel, \emph{{Matching conditions in
  metric affine gravity}},
  \href{http://dx.doi.org/10.1103/PhysRevD.66.104013}{\emph{Phys. Rev. D}
  {\bfseries 66} (2002) 104013}.

\bibitem{Sciama1963}
D.~Sciama, \emph{Recent developments in general relativity, p. 415},  1963.

\bibitem{Kibble1961}
T.~W.~B. Kibble, \emph{{Lorentz invariance and the gravitational field}},
  \href{http://dx.doi.org/10.1063/1.1703702}{\emph{J. Math. Phys.} {\bfseries
  2} (1961) 212--221}.

\bibitem{Blagojevic:2013xpa}
M.~Blagojevi\'c and F.~W. Hehl, eds., \emph{{Gauge Theories of Gravitation}: {A
  Reader with Commentaries}}.
\newblock World Scientific, Singapore, 2013,
  \href{http://dx.doi.org/10.1142/p781}{10.1142/p781}.

\bibitem{Blagojevic2001}
M.~Blagojevi\'c, \emph{{Gravitation and gauge symmetries}}.
\newblock CRC Press, 2001.

\bibitem{Obukhov2006b}
Y.~N. Obukhov, \emph{{Poincare gauge gravity: Selected topics}},
  \href{http://dx.doi.org/10.1142/S021988780600103X}{\emph{Int. J. Geom. Meth.
  Mod. Phys.} {\bfseries 3} (2006) 95--138},
  [\href{https://arxiv.org/abs/gr-qc/0601090}{{\ttfamily gr-qc/0601090}}].

\bibitem{Wald:1984rg}
R.~M. Wald, \emph{{General Relativity}}.
\newblock Chicago Univ. Pr., Chicago, USA, 1984,
  \href{http://dx.doi.org/10.7208/chicago/9780226870373.001.0001}{10.7208/chicago/9780226870373.001.0001}.

\bibitem{McCrea:1992wa}
J.~D. McCrea, \emph{{Irreducible decompositions of non-metricity, torsion,
  curvature and Bianchi identities in metric affine space-times}},
  \href{http://dx.doi.org/10.1088/0264-9381/9/2/018}{\emph{Class. Quant. Grav.}
  {\bfseries 9} (1992) 553--568}.

\bibitem{BeltranJimenez:2019hrm}
J.~Beltr\'an~Jim\'enez and F.~J. Maldonado~Torralba, \emph{{Revisiting the
  stability of quadratic Poincar\'e gauge gravity}},
  \href{http://dx.doi.org/10.1140/epjc/s10052-020-8163-8}{\emph{Eur. Phys. J.
  C} {\bfseries 80} (2020) 611},
  [\href{https://arxiv.org/abs/1910.07506}{{\ttfamily 1910.07506}}].

\bibitem{poisson2004relativist}
E.~Poisson, \emph{A relativist's toolkit: the mathematics of black-hole
  mechanics}.
\newblock Cambridge university press, 2004.

\bibitem{abramowitz1972handbook}
M.~Abramowitz and I.~A. Stegun, \emph{Handbook of Mathematical Functions with
  Formulas, Graphs, and Mathematical Tables. National Bureau of Standards
  Applied Mathematics Series 55. Tenth Printing.}
\newblock ERIC, 1972.

\bibitem{Senovilla:2014yea}
J.~M.~M. Senovilla, \emph{{Double layers in gravity theories}},
  \href{http://dx.doi.org/10.1088/1742-6596/600/1/012004}{\emph{J. Phys. Conf.
  Ser.} {\bfseries 600} (2015) 012004},
  [\href{https://arxiv.org/abs/1410.5650}{{\ttfamily 1410.5650}}].

\bibitem{xAct}
J.~M. Martin-Garcia, A.~Garc{\'\i}a-Parrado, A.~Stecchina, B.~Wardell,
  C.~Pitrou, D.~Brizuela et~al., \emph{{xAct: Efficient tensor computer algebra
  for the Wolfram Language}}, {\emph{{\tt \url{http://www.xact.es}}} (latest
  version Oct. 2021) }.

\end{thebibliography}\endgroup

\end{document}